\documentclass[referee,a4paper,12pt,traditabstract]{jswsc} 

\usepackage{graphicx}
\usepackage{txfonts}
\usepackage{subfigure}
\usepackage{epstopdf}
\usepackage[displaymath,mathlines]{lineno}
\usepackage[authoryear,round]{natbib}
\usepackage[backref]{hyperref}
\usepackage{url}
\usepackage{comment}

\hypersetup{colorlinks=true,citecolor=cyan,urlcolor=cyan,linkcolor=blue}
\DeclareUnicodeCharacter{2212}{-}
\begin{document}

%\begin{linenumbers}  

   %\title{Study of the $VB_z$ parameter variations associated with a propagating CME: a perspective for the HENON mission from EUHFORIA}

   \title{Assessing $VB_z$ variations during CME propagation: a preparatory study for the HENON mission using EUHFORIA}

%   \subtitle{Paper I}
   
   \titlerunning{EUHFORIA simulations for the HENON mission}

   \authorrunning{Prete et al.}

   \author{Prete G.
          \inst{1},
          Niemela A.
          \inst{2,3},
          Poedts S.
          \inst{4,5}
          Zimbardo G.
          \inst{1,6},
        Cicalò S.
          \inst{7},
          Marcucci M. F.
          \inst{8},
          Laurenza M.
          \inst{8},
          Stumpo M.
          \inst{8},
          Landi S.
          \inst{9},
          Sangalli M.
          \inst{9},
          Provinciali L.
          \inst{10},
          Monferrini D.
          \inst{10},
          Calcagno D.
          \inst{10},
          Di Tana V.
          \inst{10},
          Walker R.
          \inst{11},
          Pecora F.
          \inst{12},
        Nisticò G.
          \inst{1,6},
        Carbone V.
          \inst{1,6}\fnmsep\thanks{Deceased.},
         Chiappetta F.
          \inst{1}, 
        Greco A.
          \inst{1,6},
        Lepreti F.
          \inst{1,6},
          Malara F.
          \inst{1,6},      
          Perri S. 
          \inst{1,6},
          Servidio S.
          \inst{1,6}
          }

   \institute{Department of Physics, University of Calabria,
              Ponte P. Bucci, Cubo 31C, Rende, Italy\\
              \email{\href{mailto:giuseppe.prete@unical.it}{giuseppe.prete@unical.it}}
         \and
             Heliospheric Physics Laboratory, Heliophysics Division, NASA Goddard Space Flight Center, Greenbelt, MD 20771, USA
         \and
             Goddard Planetary Heliophysics Institute, University of Maryland, Baltimore County, Baltimore, MD 21250, USA
         \and
             Centre for Mathematical Plasma Astrophysics, Dept.\ of Mathematics, KU Leuven, Celestijnenlaan 200B, 3001 Leuven, Belgium 
         \and
             Institute of Physics, University of Maria Curie-Skłodowska, Pl.\ M.\ Curie-Skłodowska 5, 20-031 Lublin, Poland
         \and
             National Institute for Astrophysics, Scientific Directorate, Viale del Parco Mellini 84, I-00136 Roma, Italy
         \and 
             Space Dynamics Services S.R.L., Navacchio di Cascina,  Pisa, Italy 
         \and 
             Institute of Space Astrophysics and Planetology - INAF, Via del Fosso del Cavaliere 100, 00133, Roma, Italy
         \and 
             Dipartimento di Fisica e Astronomia, Università degli Studi di Firenze, Via G. Sansone 1, 50019 Sesto Fiorentino, Italy; INAF, Osservatorio Astrofisico di Arcetri, Largo E. Fermi 5, I-50125 Firenze, Italy; and INFN, Sezione di Firenze, Via G. Sansone 1, I-50019 Sesto Fiorentino (FI), Italy
        \and
            ARGOTEC S.R.L., San Mauro Torinese, Torino, Italy 
        \and 
            ESA-ESTEC, European Space Agency, Noordwijk, Netherlands
        \and 
            Department of Physics and Astronomy, University of Delaware, Newark, DE 19716, USA
             }

%%   \date{Received September 15, 1996; accepted March 16, 1997}

  % \abstract{}{}{}{}{}        %% uncomment if structured abstract is desired
 %% 5 {} token are mandatory
 
  \abstract
 %% context heading (optional). leave {} empty if necessary  
{Coronal mass ejections (CMEs) are among the main drivers of space weather hazards. 
In this context, HENON is a new space mission designed to carry out observations in the solar wind upstream of the Earth, aiming to provide timely alerts for hazardous perturbations propagating towards the Earth. HENON will orbit Earth on a distant retrograde orbit, approximately 0.082~AU upstream of the Earth when it is on the Sun-Earth line.
The measurements taken by HENON will allow us to determine plasma and magnetic field parameters with a lead time of several hours with respect to the Lagrangian point L1. 

We assess the $VB_z$ parameter variations (the product of solar wind speed V and southward magnetic field $B_z$) along the HENON orbit. Given its role as a primary driver of geomagnetic activity, we analyse how these measurements change with respect to Earth's position to evaluate HENON's forecasting potential.

We used the FRi3D CME model of the EUHFORIA simulation code to characterize the initial properties of the CME. FRi3D allows us to set the CME magnetic field as a magnetic flux rope.  
From the simulation results, we evaluated the $VB_z$ parameter at nine virtual spacecraft positions along the planned HENON orbit. The heliocentric longitudes of the virtual spacecraft range from about $-6.9^\circ$ to $6.9^\circ$, while 
the geocentric longitudes vary from $-60^\circ$ to $+60^\circ$ in steps of $15^\circ$. 

The initial direction of propagation of the CME central apex is either along the Sun-Earth line or at heliocentric longitudes of $\pm 30^\circ$. We find that with the proposed orbital parameters, the values of the $VB_z$ parameter along the HENON orbit are sufficiently similar to those measured in the vicinity of the Earth to be useful for space weather forecasts. At the same time, HENON permits to evaluate $VB_z$ with a lead time of about 2--8 hours, depending on the spacecraft position and the speed of the CME. The forecasting capabilities provided by HENON are expected to be foundational for the space weather community. This advancement has direct implications for enhancing the resilience of satellite communications and safeguarding critical infrastructure against space weather events.

   }        %% replace by pair of curly brackets, {}, if structured abstract is selected


   \maketitle
%%
%%________________________________________________________________

\section{Introduction}

The availability of early Space Weather (SW) warnings is of paramount importance for taking effective mitigation measures against space weather hazards \citep{echer2005introduction,gopalswamy2022SW}. As is the case with forecasts of conventional weather, extended lead times and accurate predictions significantly improve our ability to reduce risks and minimize damage \citep{macalester2014extreme,FRY2012}. Due to the diversity in the scales of the phenomena that influence and drive SW, this task is challenging. 

Our current capabilities include (but are not limited to): remote sensing and imaging of the solar corona and photosphere in many different wavelengths, which target early identification of strong energy release sites \citep{brueckner1995large,kaiser2008stereo,lemen2012atmospheric,vourlidas2016wide,antonucci2020metis}; real-time in-situ monitors upstream of Earth (at the  Lagrange point L1) \citep{Wind1995,SOHO1995,ACE1998} and close to Earth at geosynchronous orbit \citep{Hill2003,Iyer2006JApA,Lotaniu2023,Stiefel2025A&A}; missions orbiting the Sun that provide different perspective of these phenomena \citep{Kaiser2005AdSpR, muller2020A&A, Fox2016}.  
Previous missions, such as STEREO \citep{webb2010using, ravishankar2019estimation}, have transited the Lagrangian point L4 to monitor solar activity and space weather. Future missions are planned for both L4 and L5 \citep{gopalswamy2011earth,rodriguez2020space,posner2021multi}, including Vigil, which will focus on improving the SW forecasting \citep[e.g.,][]{Eastwood24}. 
Furthermore, the recent PUNCH mission \citep{deforest2026polarimeter}, which consists of 4 imagers in a Sun-synchronous orbit around the Earth, will provide an unprecedented point of view on SW events. 

In a comprehensive review, \citet{Vourlidas19} suggested that an upstream monitor at 0.3~AU from Earth would be necessary to provide warnings of solar wind perturbations with approximately 24-hour advance notice. However, 
Kepler's laws do not allow a spacecraft to stay stationarily on the Sun-Earth line at 0.3~AU from the Earth on a free-fall (Keplerian two-body) orbit. 
As a first step in the direction upstream of the Earth, we can place a spacecraft at the collinear Lagrange point L1 of the Sun–Earth system, which lies at  a much smaller distance, approximately 0.01~AU from Earth toward the Sun, than suggested by \cite{Vourlidas19}.

\begin{figure}[!h]
\centering
\includegraphics[width=1\textwidth]{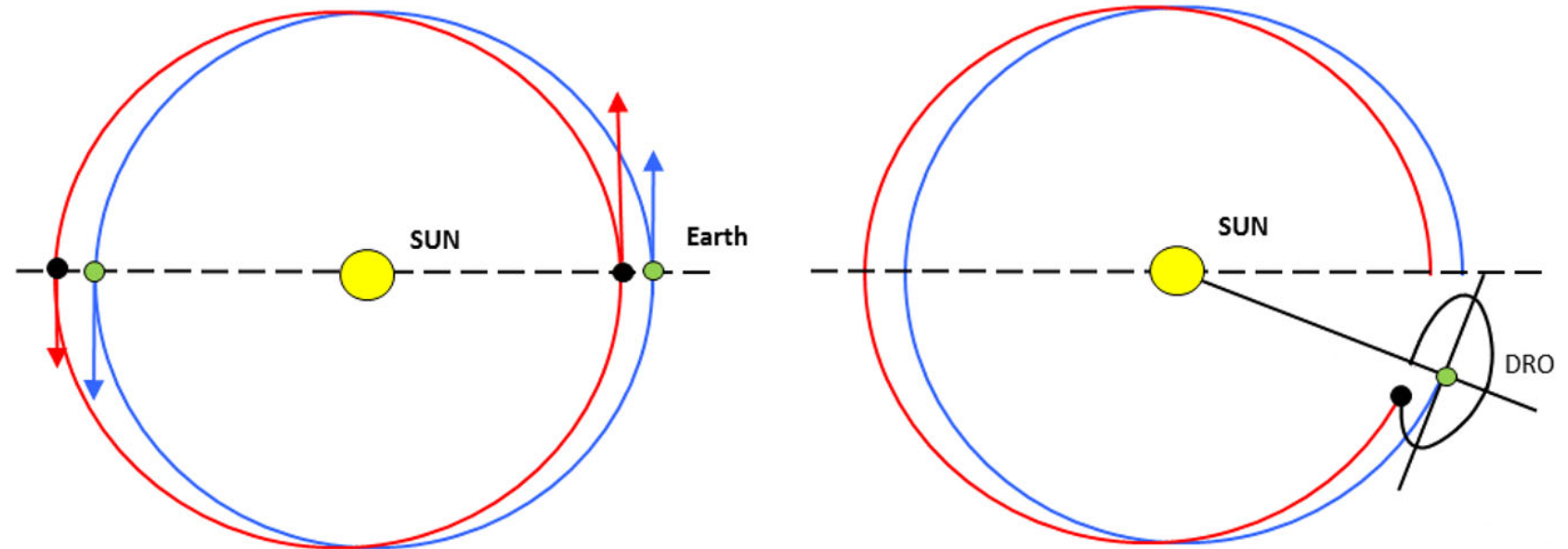}
\caption{Dynamical configuration of DRO in the ecliptic plane, from \citet{Perozzi17}. In both panels, the heliocentric orbit of the Earth is depicted in blue, while the orbit of the satellite is shown in red. In the right panel, the spacecraft's trajectory with respect to the Earth's reference frame is shown in black. }
\label{fig0}
\end{figure}

To go further upstream, in the planar circular restricted three-body problem, it is possible to take advantage of the specific type of orbits called Distant Retrograde Orbits (DROs), which were studied in detail by M. Hénon using Hill’s approximation \citep{Henon69,Henon70}. 
In such a case, both the spacecraft and the Earth are orbiting around the Sun, and, because of the same semi-major axis (but different eccentricities), they have the same revolution period around the Sun.  
When viewed in the synodic reference frame (a rotating coordinate system commonly used for multi-body orbital dynamics), the spacecraft trajectories display apparent retrograde motion relative to the planet, as shown in Fig.~\ref{fig0}. 

The DRO approach is the theoretical framework that forms the basis of the HEliospheric pioNeer for sOlar and interplanetary threats defeNce (HENON) mission. The ESA-ASI HENON mission \citep{Provinciali24,Cicalo25} aims to send a spacecraft on a DRO orbit to test the feasibility of the aforementioned approach. The HENON mission payload includes instruments to measure the solar wind plasma, magnetic field, and fluxes of energetic ions and electrons. The planned upstream distance of the spacecraft is about 0.082$\sim$AU, which allows a 2--8 hours advance notice with respect to a spacecraft located in L1. For the HENON mission, the period of revolution around the Sun is 1 year (semi-major axis of 1 AU), and, depending on its eccentricity, the nominal orbit is chosen such that the spacecraft remains at a much larger distance from the Earth than the Lagrangian point L1 for an extended period. This orbital configuration offers highly favourable conditions for real-time SW monitoring, while also ensuring stable relative orbital motion. 
Such a DRO allows a single spacecraft to dwell upstream of the Earth for only a limited time. Thus, a multispacecraft mission will be necessary to have continuous monitoring of solar wind conditions upstream of the Earth. 
A similar concept for a SW mission, the Diamond mission, \citep{Cyr00}, envisioned a constellation of four spacecraft equally spaced along the same DRO at a minimum distance of 0.1~AU from Earth, but has never been realised. Possible exploitation of DROs for asteroid impact monitoring has also been proposed \citep{Perozzi17}. 

The possibility of sending space weather monitors upstream of L1, that is, of doing sub-L1 measurements, has been considered by several authors \citep[e.g.,][]{Lugaz24b,Lugaz25,Palmerio25}. However, a crucial aspect to evaluate during the mission's planning phase is the correlation between observations acquired by the sub-L1 spacecraft and those obtained at Earth, given that the latter are more directly linked to SW disturbances. Indeed, it is well known that the interplanetary medium is characterised by complex physical phenomena, including shocks, turbulence, current sheets, and coherent structures \citep{brunocarbone2005review, matthaeus2015intermittency}. These phenomena induce spatial and temporal variability in the properties of solar wind and energetic particles. For instance, \citet{Lugaz18} found, by analyzing Wind and ACE data for a number of CMEs, that the magnetic field components at the two spacecraft are very well correlated when the longitudinal separation is less than 0.008 AU, which corresponds to the correlation length of the magnetic field at 1 AU \citep{ruiz2014characterization, cuesta2022isotropization}, but the correlation decreases when the separation increases. Also, \citet{Scolini24} analyzed the coherence of Alfvénic fluctuations within CMEs observed by ACE and Wind at longitudinal separation of 0.009 - 0.013 AU, and found that the magnetic field profile seen by ACE and Wind is well correlated within the magnetic ejecta, but less so in the sheaths.  However, we can argue that larger-scale structures, which are more geoeffective, have larger scale coherence than Alfvenic fluctuations.

At the same time, CMEs evolve with radial distance, which also limits the radial separation between a sub-L1 monitor and the Earth. Using Solar Orbiter \citep{muller2020A&A} as an upstream monitor at 0.5 AU, \citet{Laker24} compared the observations for two CMEs, finding that in one case the time profile of $B_z$ was in good agreement between Solar Orbiter and Wind, despite of the radial distance and a longitudinal separation of about 10°. Regarding the second CME event, \citet{Laker24} found that the magnetic structure was typical of two interacting flux ropes, but the agreement of the $B_z$ time profile between spacecraft was poorer. Considering the strong geomagnetic storm of 10-12 May 2024, when the STEREO-A spacecraft was situated at 0.956 AU and 12.6° west of Earth, \citet{Weiler25} find that the shock was observed 2.57 hr earlier at STEREO-A than at L1, which is consistent with the shock speed of about 700 km/s. In addition, the $B_z$ profiles showed good agreement, and the intensity of the geomagnetic storm was predicted well using STEREO-A data. 

In another study, \citet{Lugaz24a} analysed the structure of magnetic ejecta in CMEs  using STEREO-A and Wind data, finding that the typical angular size of magnetic ejecta is 20$^\circ$--30$^\circ$ for most events, which is smaller than typically assumed. This implies that a sub-L1 monitor cannot stay too far away from the Sun-Earth line. In a recent, comprehensive study on the properties required for a sub-L1 mission, \citet{Lugaz25} suggest that a longitudinal separation from the sub-L1 monitor to the Sun-Earth line of $< \sim 15^\circ$ is needed, with a separation of less than $10^\circ$ to be preferred to ensure that CMEs measured by the sub-L1 monitor also impact the geospace environment. 

HENON orbit corresponds to 0.082 AU upstream of the Earth and to a maximum distance of $\pm 0.164$ AU along the $y$ direction of the GSE system \citep{Cicalo25}. In such a case, the maximum angular separation of HENON with the Sun-Earth line is $\pm 9.3^\circ$. Furthermore, the prime region for space weather measurements, i.e., KR1, see Fig. \ref{fig1}, is defined as $\pm 60^\circ$ in geocentric longitudes to stay upstream of the Earth: this corresponds to $\pm 6.9^\circ$ in heliocentric longitude. Therefore, HENON KR1 orbit is well within the angular separation recommended by \citet{Lugaz25} for carrying out useful measurements.

This paper investigates the relationship between observations from a virtual spacecraft situated on a DRO and those recorded at Earth. To achieve this, we employ a large-scale magnetohydrodynamic (MHD) simulation of the heliospheric structure, using the EUHFORIA code \citep[EUropean Heliospheric FORecast Information Asset; ][]{Pomoell18}. Additionally, we incorporate the orbital ephemerides of the operational DRO, which were numerically computed by SpaceDyS during the Phase A and B mission analysis studies for HENON \citep{Provinciali24,Cicalo25}. It should be noted that numerical simulations have a finite spatial resolution and that each model makes several assumptions, such as those regarding the coronal structure, the initial magnetic field, and the simulation domain. Global MHD simulations of the heliosphere are not yet able to resolve the small-scale structures of CMEs, and this should be taken into account when interpreting the simulation results; nevertheless, such simulations can provide very valuable information on the large scales.

We initialized EUHFORIA with a uniform background solar wind in which the CME is propagating. We compared the simulation results at HENON, located at various positions along the DRO, and at the Earth. We performed several runs, varying the magnetic helicity of the initial magnetic flux rope that describes the CME magnetic field, as well as the direction of propagation of the CME relative to the Sun-Earth line, to study various scenarios for the interaction of CMEs with our planet. We use the FRi3D model \citep{Maharana2022} of EUHFORIA to simulate the evolution of the CME. Among the different quantities of interest for SW forecasts, we concentrate on $VB_z$, where $V$ is the radial component of the solar wind velocity and $B_z$ is the vertical component of the interplanetary magnetic field (IMF) in the Geocentric Solar Magnetospheric (GSM) coordinate system. It is well-known that the parameter $VB_z$ is an estimator for the reconnection electric field located at the nose of the Earth’s magnetosphere. In this regard, $VB_z$ demonstrates notable correlation with the onset of geomagnetic storms \citep{Gonzalez89,Gonzalez94,Wang03,Spencer11}. However, it is necessary to underline that $VB_z$ remains only a proxy for the reconnection rate, considering that the electric field characteristic of the solar wind can differ from the field observed at the magnetopause \citep{Borovsky14}.

Here, we focus on $VB_z$ both because of its significance for geomagnetic storms and also because it is a typical parameter whose value can be evaluated at HENON and compared with measurements near Earth.

\section{Geomagnetic storms, EUHFORIA Model and Numerical Setup}

\subsection{Rationale}

The product $VB_z$ serves as a predictor for geomagnetic storm indices such as $Dst$, which is a measure of the intensity of the ring current around the Earth \citep{Mayaud1980GMS}, and SYM-H that quantifies the strength of a magnetic storm \citep{Latiff2024}. Numerous studies have established that intense geomagnetic storms (defined as $Dst < -100\;$nT) are highly probable when the magnitude of the southward IMF component, $|B_z|$, exceeds $6$--$10\;$nT and the convection electric field, $VB_z$, is less than $-5\;$mV/m for a duration exceeding 3 hours \citep{Gonzalez89,Gonzalez94,Wang03,Spencer11,Verbanac13}. The threshold of $-5\;$mV/m can be obtained, as an example, considering the solar wind plasma propagating at 500 km s$^{-1}$ and carrying a southward magnetic field of 10 nT, i.e., $E_y=V_xB_z=5\times 10^5$ m s$^{-1} \times (- 10^{-8})$ T = $-5\times 10^{-3}$ V/m. 

Consequently, a measurement period of at least 3 hours in the solar wind is required following the arrival of an ICME at HENON before a geomagnetic storm alert can be reliably issued. For instance, to provide a 6-hour advance alert, observations should begin 6 hours before the southward turning of the magnetic field at Earth. Geoeffective ICME speeds near 1~AU typically range from $400$--$2000\;$km/s, with faster ICMEs generally correlating with more severe storms \citep{Liu14,Temmer15}. 
Here, we assume an average geoeffective ICME speed of $700\;$km/s as a typical speed of the more common events. (Clearly, extreme events with large speeds are of the highest concern for space weather, and we reserve the study of larger speeds for the future.) A 6-hour advance alert requires the sentinel spacecraft to be positioned at an upstream distance of approximately $15,120,000\;$km from Earth $(\sim 0.1\;$AU). It is important to note that the lead time is reduced for CMEs faster than 700 km s$^{-1}$. Nonetheless, it is crucial to recognize that even relatively slow CMEs can exhibit significant geoeffectiveness, as highlighted by \citet{Gopalswamy22}.

To define efficient alert algorithms for the prediction of geomagnetic storms, we need to assess the
difference in the observed $VB_z$ at the Earth and along the HENON orbit using numerical simulations. We have been using two well-known global simulation models for the propagation of perturbations in the heliosphere, namely ENLIL \citep{Odstrcil03} and EUHFORIA \citep{Pomoell18}. Here, we present the results obtained with EUHFORIA. In the following simulations, we consider the evolution of a hypothetical CME, allowing us to select specific features and the positions of the observing spacecraft. Since the simulation epoch is arbitrary and not related to the inclination of the Earth's magnetic dipole and the Earth rotation, we will refer to the IMF component perpendicular to the ecliptic, namely $B_z$ in the Geocentric Solar Ecliptic (GSE) coordinate system, and not to the actual southward component $B_S$ in the Geocentric Solar Magnetospheric (GSM) system, which depends on time according to the Earth daily rotation and position along its heliocentric orbit. In any case, $B_z$ is a good estimator of $B_S$ \citep{Wang03}. 

\subsection{EUHFORIA model}

The EUHFORIA model, developed by \citet{Pomoell18}, is a space weather modelling tool that computes the time evolution of the inner heliospheric plasma environment using a combination of empirical and physics-based modelling approaches. EUHFORIA consists of two major components: a coronal model and a heliospheric
model. The data-driven coronal model uses synoptic magnetograms as input and provides solar wind plasma parameters at 0.1~AU from the Sun, utilising the empirical Wang-Sheeley-Arge model \citep{schatten1969model, altschuler1969magnetic,Arge2003AIPC}.
%The coronal model provides data-driven solar wind plasma parameters at 0.1 AU by constructing a magnetic field model of the coronal large-scale magnetic field and employing empirical relations to determine the plasma state such as the solar wind speed and mass density.
These are then used as boundary conditions to drive a 3D time-dependent MHD model of the inner heliosphere up to 2~AU. Different modelling approaches are available in EUHFORIA to simulate CME evolution, including the cone model \citep{Xie2004JGRA}, in which the CME has a spherical plasma bubble shape with a uniform density and no magnetic field structure; the spheromak model \citep{Kataoka2009JGRA, verbeke2019A&A} that assumes a spherical shape for the CMEs with flux-rope structures; and the FRi3D \citep[Flux Rope in 3 Dimensions, ][]{Isavnin_2016, Maharana2022} model , which can reproduce the CMEs shape and its deformations with a flux-rope internal structure. Previous runs (not shown) assumed the CME as a force-free spheromak, which gives rise, during CME propagation, to a magnetic flux rope structure  \citep{Chandrasekhar1957ApJ,Shiota2016,Scolini2019,verbeke2019A&A} . 

The runs presented here were carried out with the latest FRi3D implementation in EUHFORIA \citep{Maharana2022}. In this novel version, the initial CME is described as a flux rope, according to the Graduated Cylindrical Shell \citep[GCS; ][]{Thernisien06,Thernisien11} model. The GCS model is an empirical model widely used by the solar physics community for describing CMEs.  
The FRi3D model is an analytical model that provides the geometric shape of the CME and its deformations. It reproduces a cylindrical shell model filled with magnetic field lines. A series of parameters, such as the pancaking, i.e. the deformation of the frontal part of a flux rope during the expansion in the poloidal and toroidal direction, the flattening, which take into account for the compression of the front part of the flux rope, and the rotational skew, namely the flux rope deformation due to the solar rotation, permit us to model the global shape of the CME accurately. The {\it Lundquist model} \citep{Lundiquist1950} is used to characterize the helical and twisted field lines in the force-free magnetic field distribution, with the aim of determining the magnetic field strength in a cylindrical geometry \citep[for more details, see ][]{Maharana2022}. A series of parameters has been set to shape the flux-rope inside the cylindrical shell: the angular half-width ($\varphi_{hw}$), i.e., the maximum angular extension in the azimuthal direction, the angular half-height ($\varphi_{hh}$), namely the maximum angular extension in the polar direction, the toroidal height ($R_t$) and the poloidal height ($R_p$), which indicate, respectively, the heliocentric distance to the apex of the CME axis and the radius of the cross-section at the apex of the CME. The total speed of the CME is given by the contribution of height growth due to linear propagation and the increase in the radial cross section due to the pancaking effect $v_{3D}=\frac{d}{dt}(R_t+R_p)=v_{R_t}+v_{R_p}$.
The magnetic field is characterized by five parameters: the tilt, i.e., the angular orientation of the CME axis measured from equatorial plane, the magnetic flux, namely the total magnetic flux of the CME, the twist, i.e., the number of turns in the magnetic field, the chirality that indicates the handedness of the flux-rope, and the polarity that gives the direction of the axial magnetic field of the flux-rope. Finally, the cylindrical shell is filled with a uniform-density plasma \citep[see Table~1 in ][]{Maharana2022}.

\subsection{Simulation set-up}

We conducted several runs, varying parameters such as the CME initial speed, the CME heliographic longitude, and the flux rope magnetic helicity sign. The solar wind conditions before the CME propagation have been set as 
uniform, with n = 5.35 cm$^{-3}$, V = 472 km/s and $|B_r|$ = 2.10 nT, to flatten the influence of coronal structures other than the CME.
The use of EUHFORIA is advantageous compared to other heliospheric models because we can set the position of the observing spacecraft, and we can fix the CME's initial magnetic field properties. In addition, the high spatial resolution of EUHFORIA ($\Delta r$ = 0.004 AU, $\Delta \theta$ = 2$^{\circ}$, $\Delta \phi$ = 2$^{\circ}$) enables the distinction, in the simulation results, between the formation of the CME-driven shock wave and the shocked sheath region preceding the magnetic cloud of the CME. 

%_____________________________________________________________
%                 A figure as large as the width of the column
%-------------------------------------------------------------
   \begin{figure}
   \centering
   \includegraphics[width=0.9\textwidth]{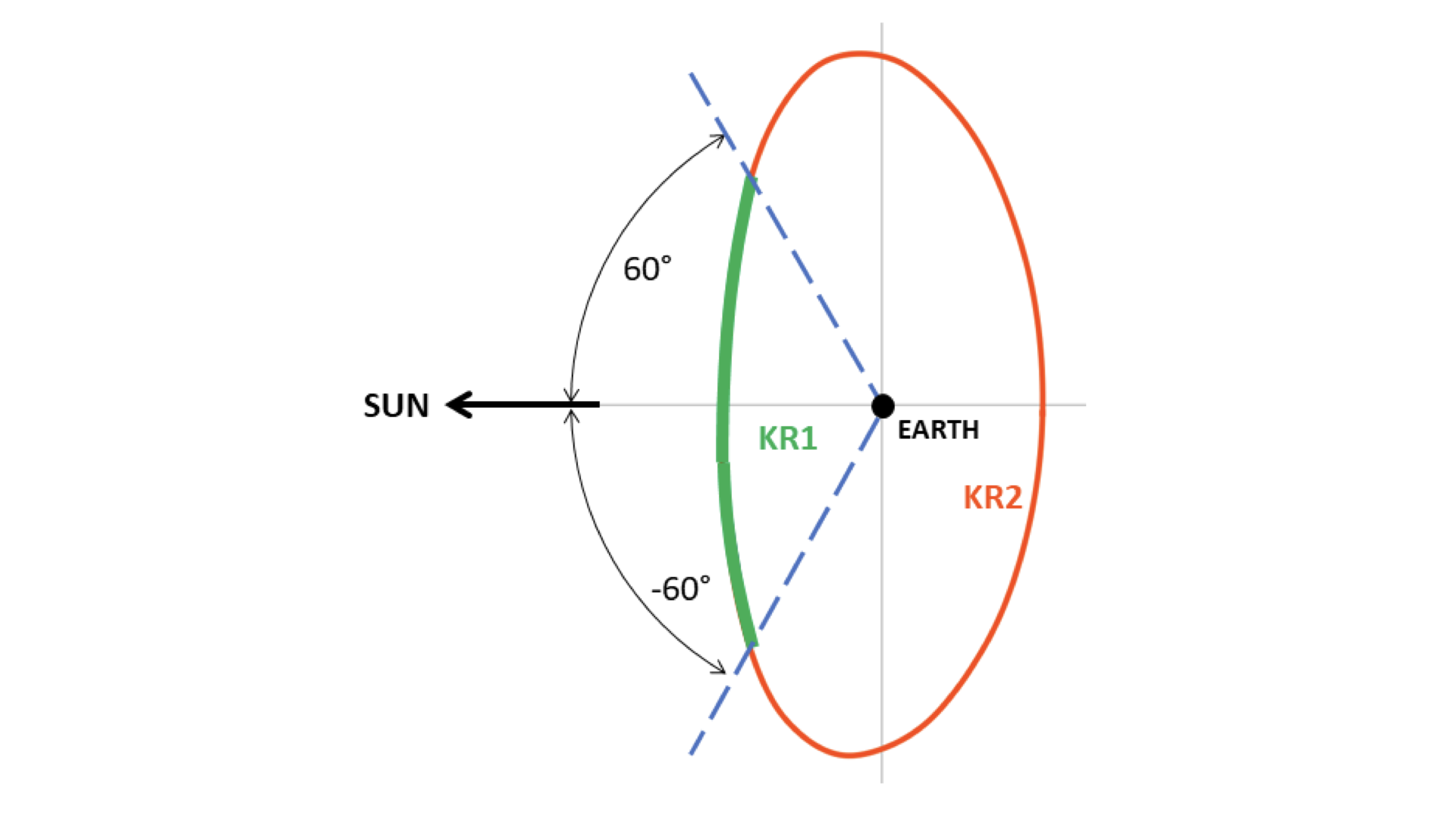}
      \caption{The HENON spacecraft orbits around the Earth. Key region 1 (KR1) and key region 2 (KR2) are also reported. The spacecraft geocentric longitudes are measured from the Earth-Sun line in the clockwise direction, in agreement with the retrograde motion of HENON.
      (Figure adapted from \cite{Cicalo25})
              }
         \label{fig1}
   \end{figure}

Furthermore, the orbital ephemerides of the HENON trajectory computed by SpaceDyS have been exploited to simulate the acquisition of measurements at nine distinct positions along the DRO within the key region 1 (KR1; see Fig.~\ref{fig1}). Here, KR1 is defined as the region upstream of the Earth, which has a geocentric angular width of $\pm 60^\circ$ with respect to the Sun-Earth line.  Looking from the Sun, the HENON longitudes in KR1 go from $-6.9^\circ$ to $6.9^\circ$.
Typically, HENON will spend 90 days in KR1 and the rest of the year in key region 2 (KR2). For the present runs, the minimum upstream distance of HENON is $0.082\;$AU \citep{Cicalo25}. 
To understand how the $VB_z$ parameter would change along the DRO orbit and with respect to Earth, we have inserted nine virtual spacecraft, placed in KR1, spaced by $15^\circ$ in terrestrial longitude, that cover the longitude range $\pm 60^\circ$ (as shown in Fig. \ref{fig1}) where the properties of the CME and the associated shock wave, obtained by the simulation, will be ``monitored''. The corresponding heliocentric longitudes are $0^\circ$,  $\pm 1.4^\circ$, $\pm 2.9^\circ$, $\pm 4.7^\circ$, and $\pm 6.9^\circ$.

\begin{table}[!h]
\centering
%\resizebox{\textwidth}{!}{%
\begin{tabular}{l c}
\hline
\textbf{Parameter} & \textbf{Value} \\
\hline
$\Delta$T (hours) since the insertion time of the CME & 39\\
$V_{CME}$ (km/s) & 750 \\
R$_t$ (R$_{\odot}$) & 16.5 \\
Half height ($^{\circ}$)& 15 \\
Tilt angle ($^{\circ}$) & 0.5 \\
Flattening & 0.5 \\
Pancaking & 0.5 \\
Twist & 1 \\
Polarity & +1 \\
Flux (Wb) & $10^{13}$ \\
Mass density (kg cm$^{-3}$) & $10^{-17}$ \\
Temperature (K) & $8 \times 10^{5}$ \\
\hline
\end{tabular}%
%}
\caption{CME parameters used for all the simulations.}
\label{tab:Table1}
\end{table}

The simulation runs have been initialized with a CME speed of $750\;$km/s, with a mass density of 10$^{-17}\;$kg cm$^{-3}$ and a temperature of $8 \times 10^{5}\;$K. The time since CME insertion, $\Delta T$, is 39 
hours. The parameters which are kept constant for all simulations are presented in Table \ref{tab:Table1}. 
The parameters that were varied are presented in Table \ref{tab:Table2}. In particular, the central latitude of the CME apex was set to zero, the central longitude of the CME apex was fixed either to zero or to $\pm 30^\circ$, with a half-angle (the half-width) between the CME legs of 30 degrees, a tilt angle of 0.5 degrees (that is, the CME croissant lies in the solar equatorial plane), and the magnetic helicity was set to $+1$, for left-handed chirality, and $-1$ for right-handed chirality \citep{Maharana2022}. The FRi3D chirality is implemented, which uses an opposite convention with respect to the usual one. Clearly, many other physically relevant choices can be made for the simulation parameters, and we consider the present runs as a first study.

\begin{table}[ht]
\centering
%\resizebox{\textwidth}{!}
{
\begin{tabular}{c c c c c c c }
\hline
  & \textbf{RUN1} & \textbf{RUN2} & \textbf{RUN3} & \textbf{RUN4} & \textbf{RUN5} & \textbf{RUN6}\\
\hline
Lat ($^{\circ}$) & 0 & 0 & 0 & 0 & 0 & 0  \\
\hline
Lon ($^{\circ}$) & 0 & 0 & 30 & 30 & -30 & -30  \\
\hline
Helicity & 1 & -1 & 1 & -1 & 1 & -1 \\
\hline

\end{tabular}%
}
\caption{CME parameters used for the runs with the FRi3D model of EUHFORIA.}

\label{tab:Table2}
\end{table}

\section{Simulation Results}

%_____________________________________________________________
%                 A figure as large as the width of the column
%-------------------------------------------------------------
   \begin{figure}
   \centering
   \includegraphics[width=1\textwidth]{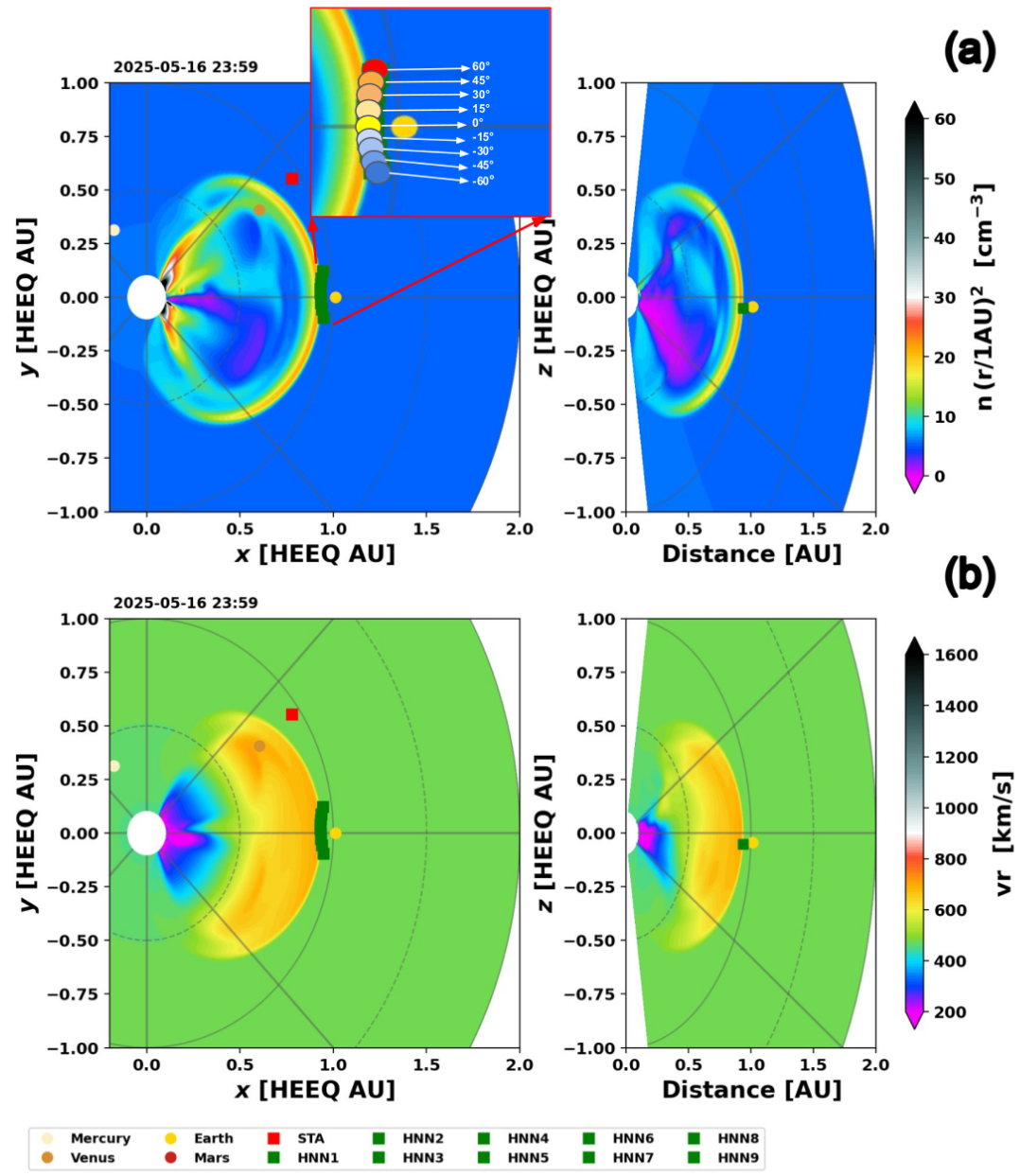}
      \caption{RUN1 simulation results at the time of arrival of the CME-driven shock at the HENON virtual spacecraft, indicated as HNN (see legend), positioned directly in front of the Earth, along the Sun-Earth line. Top row: detrended density in the equatorial plane (left) and in the meridional plane (right), with the colorbar on the right-hand side (panel (a)). Bottom row: radial velocity in the equatorial plane (left) and in the meridional plane (right), with the colorbar on the right-hand side (panel (b)). A yellow dot indicates the Earth's position, while green squares indicate the virtual spacecraft locations. In the inset we indicated the geocentric position of the virtual spacecraft in different colors.}
         \label{fig2}
   \end{figure}
%

%_____________________________________________________________
%                 A figure as large as the width of the column
%-------------------------------------------------------------
   \begin{figure}
   \centering
   \includegraphics[width=1\textwidth]{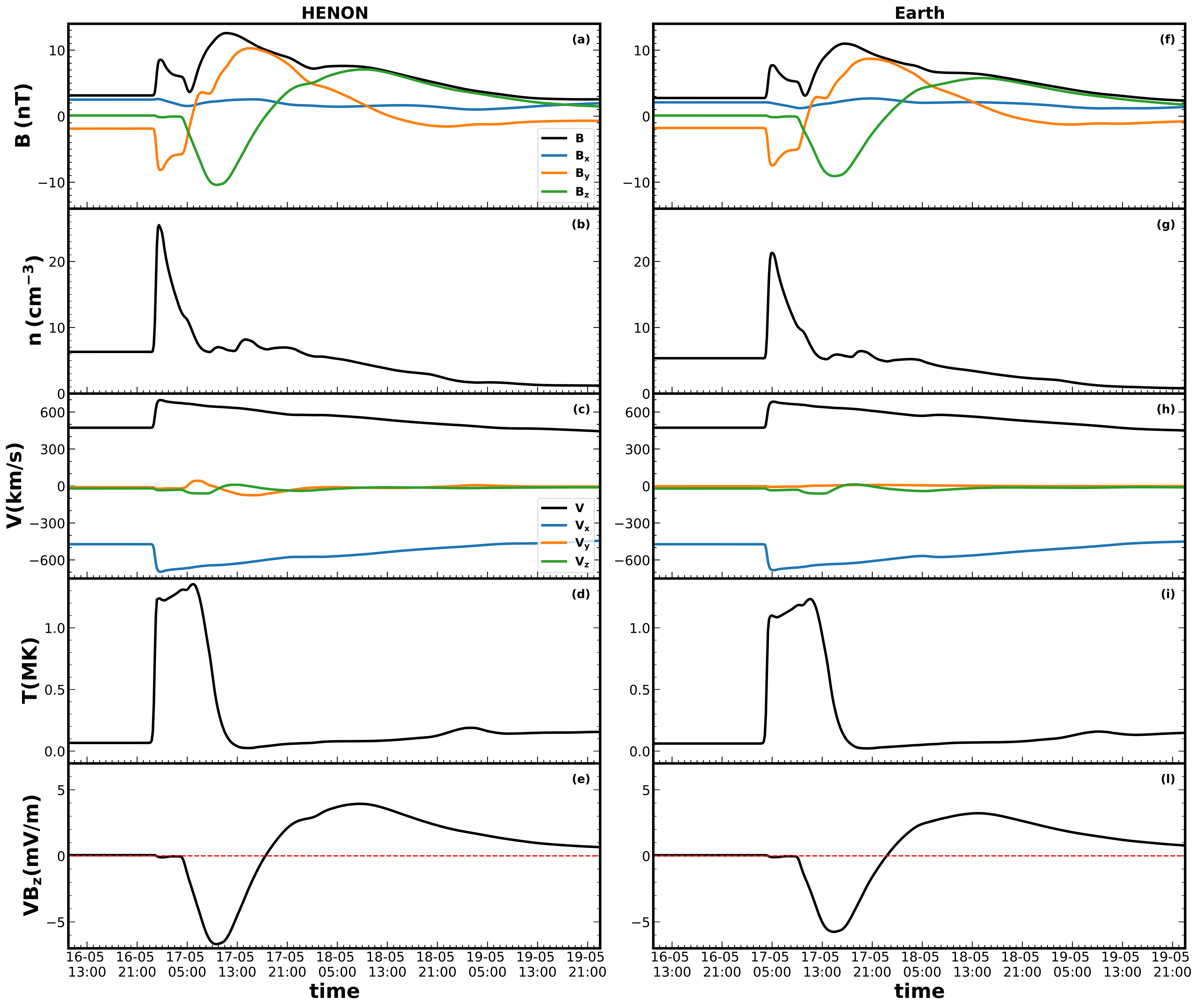}
      \caption{RUN1 simulation results. From top to bottom, the magnetic field components and strength, the plasma density, the bulk velocity components, the plasma temperature, and the $VB_z$ parameter. Left panels: values at the HENON spacecraft upstream of the Earth on the Earth-Sun line; right panels: values at the Earth.
              }
         \label{fig3}
   \end{figure}

Figure~\ref{fig2} shows the simulation for RUN1. The upper panels show the density, rescaled by a factor $(r/1 ~AU)^{2}$, and the lower panels show the radial velocity. For each quantity, the left panels show an equatorial cross-section, and the right panels show a meridional cross-section passing through the Earth, which is indicated by a yellow dot. The positions of the planets and the HENON spacecraft on the equatorial plane are also indicated as green squares.
At approximately 0.082~AU, the CME-driven shock exhibits a clear density enhancement ($\sim$20 cm$^{-3}$) relative to the ambient solar wind ($\sim$4 cm$^{-3}$) (see panel (a)), along with a significantly higher speed of  $\sim$ 700 km/s compared to the background solar wind velocity $\sim$ 500 km/s (see panel(b)).

Figure \ref{fig3} shows the time profiles of several quantities measured by the virtual HENON spacecraft at 0.082~AU upstream of the Earth, and at the Earth. The panels, from top to bottom, display the magnetic field components and magnitude, plasma number density, velocity components and magnitude, temperature, and $VB_z$. 
At $\Delta$T=39 hours from the CME injection, each quantity exhibits a jump due to the CME-driven shock crossing. Just downstream of the shock, it is possible to recognize the sheath region characterized by a more disordered magnetic field, where the temperature increases abruptly. Past the sheath, the magnetic obstacle (the core of the helical flux tube inside the FRi3D) is located, where the magnetic field components $B_y$ and $B_z$ are observed to show the typical smooth rotation \citep{gosling1990coronal}, while the $B_x$ component is almost zero. 
The $VB_z$ time profile has a typical behaviour for many geomagnetic storms \citep{Wang03}, featuring small fluctuations around zero at the shock, followed by a gradual decrease of $VB_z$, down to values less than $-5\;$mV/m, i.e., less than the threshold usually adopted for geomagnetic storm prediction for times longer than 3 hours \citep{Gonzalez1987}.
The perturbation is first detected at HENON approximately 6 hours before it is observed at Earth, consistent with the kinematic propagation of the ICMEs.

   \begin{figure}[!h]
   \centering
   \includegraphics[width=0.9\textwidth]{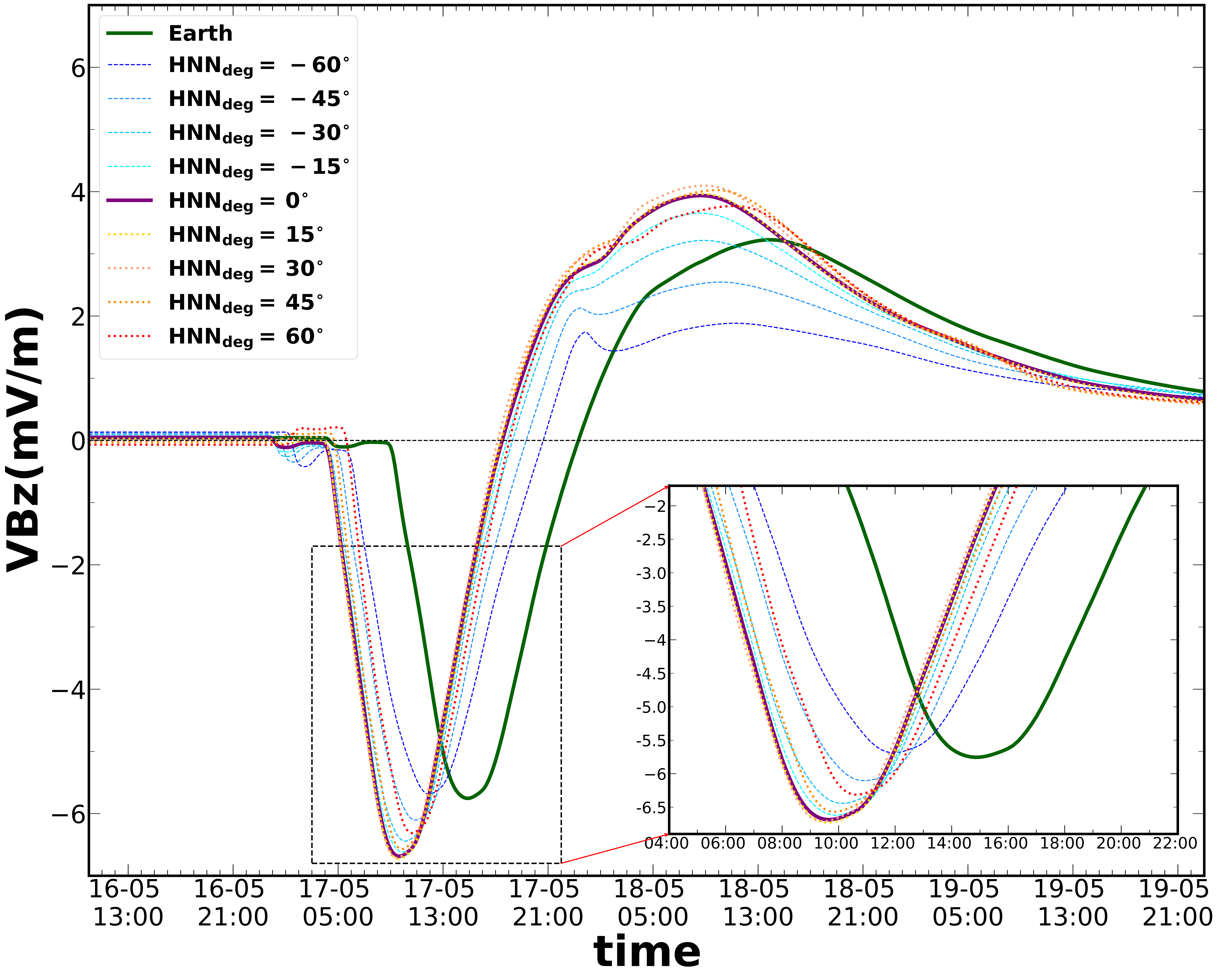}
      \caption{Variation of the $VB_z$ parameter at the nine virtual spacecraft along KR1, plus at the Earth (solid green line), in the case of positive magnetic helicity. The geocentric longitudes of each spacecraft are indicated in the legend. 
      $VB_z$ is negative just after the shock crossing and then, because of the rotation of the magnetic field inside the flux rope, positive. In this case, $VB_z$ becomes smaller than $-5\;$mV/m for a period longer than 4 hours, implying that a geomagnetic storm is possible. For later times, after the CME passage, $VB_z$ tends towards small values, typical of unperturbed periods. The inset shows a magnified view of the $VB_z$ profiles around the time of the negative peaks, as indicated by the dashed rectangle in the main panel. 
              }
         \label{fig5}
   \end{figure}

Figure~\ref{fig5} shows the simulated time series of all HENON virtual spacecraft and at the Earth for RUN1. In Fig.~\ref{fig5} we use the following graphic conventions: the value of $VB_z$ at the Earth is indicated by a solid green line, those at HENON, just in front of the Earth, by the solid purple line, those westward of the Earth-Sun line (i.e., with positive geocentric angles) by dotted lines in the shades of red, and those eastward of the Earth-Sun line (i.e., with negative geocentric angles) by dashed lines in shades of blue. A zoom on the negative $VB_z$ peaks is shown in the inset of Fig.~\ref{fig5} to better highlight the differences between the lines corresponding to the various virtual spacecraft. 

As a general remark, we can see that the overall behaviour of $VB_z$ is similar at the Earth and at HENON, with little to no dependence on the longitudinal position of the virtual spacecraft within KR1. It can be seen that negative peak values are almost all larger at HENON, as compared to the Earth: this can be understood as the behaviour of the transverse components of the interplanetary magnetic field, which scale with the heliocentric distance as $B_z \sim 1/r$, in agreement with the Parker spiral model and as observed in a recent survey of the IMF \citep{maruca2023trans}. With a positive helicity of the magnetic flux rope, the local magnetic field $B_z$ is first negative and then positive. The negative values of $VB_z$ reach $-5$~mV/m for more than three hours, indicating that a geomagnetic storm is expected. Thereafter, a prolonged period of positive $VB_z$ is observed; this is due to the spacecraft being crossed by the magnetic axis of the CME magnetic flux rope, beyond which $B_z$ turns from southward ($B_z < 0$)  to northward ($B_z > 0$). The lead time is about $5.5$ hours when HENON is on the Earth-Sun line; one can see that this lead time decreases to about $3.5$ hours when HENON is at $-60^\circ$ (blue dashed line).

\begin{figure}[!b]
   \centering
   \includegraphics[width=1\textwidth]{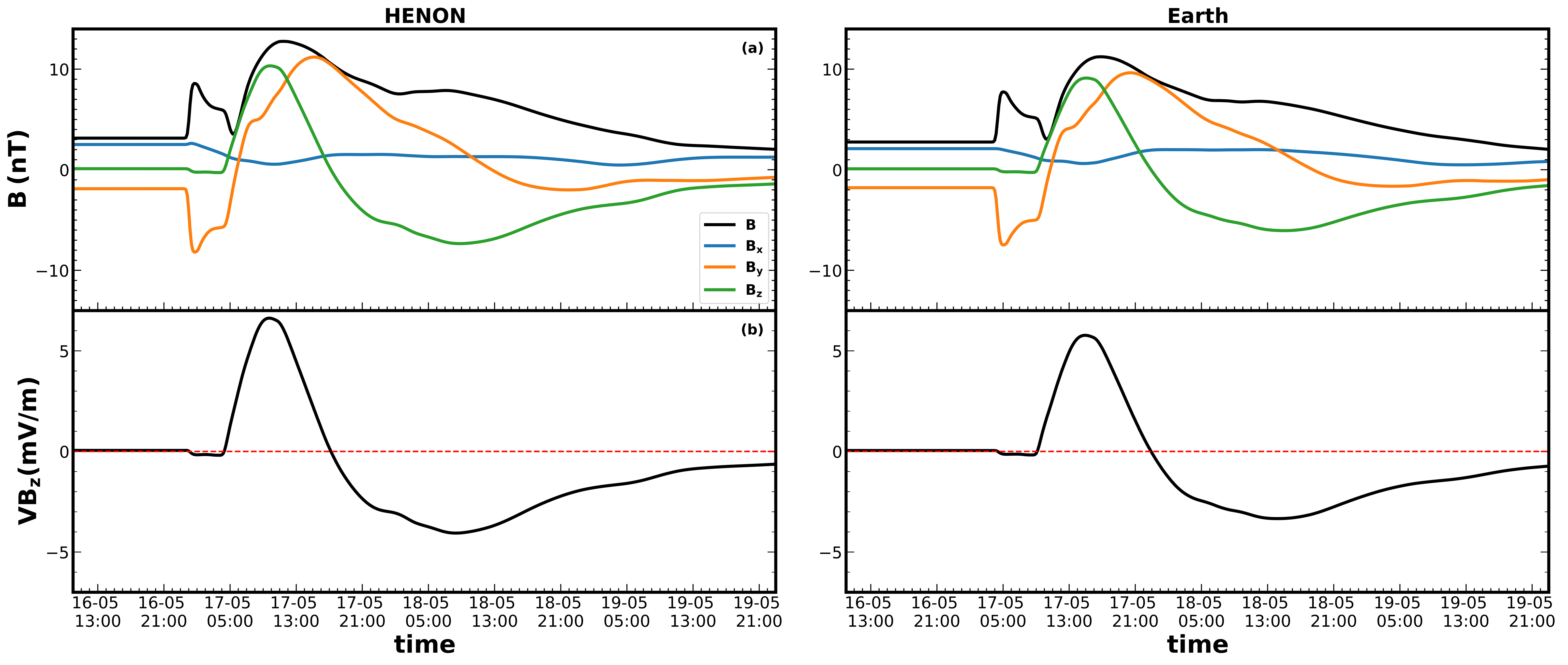}
      \caption{Same as Fig.\ref{fig3}, but for the case of a magnetic flux rope with negative magnetic helicity (RUN2). The magnetic field rotates in the opposite direction, as compared to Fig.\ref{fig3}.
              }
         \label{fig4}
\end{figure}

We consider that this decrease in lead time is due to the curved shape of the spacecraft trajectory, as shown in Fig.~\ref{fig1}, as well as the curved CME front \citep[e.g.,][]{Shen14}. Interestingly, the decrease in lead time is not symmetric for the eastward and westward positions of HENON, as shown in Fig.~\ref{fig5}. This may be related to the  deflection in the propagation of a CME when it interacts with the spiral magnetic field of the solar wind. As shown by \citet{wang2004deflection}, CMEs faster than the ambient solar wind are deflected eastward (looking from the Earth), while slower CMEs are deflected westward.

%_____________________________________________________________
%                 A figure as large as the width of the column
%-------------------------------------------------------------
   \begin{figure}[!h]
   \centering
   \includegraphics[width=0.9\textwidth]{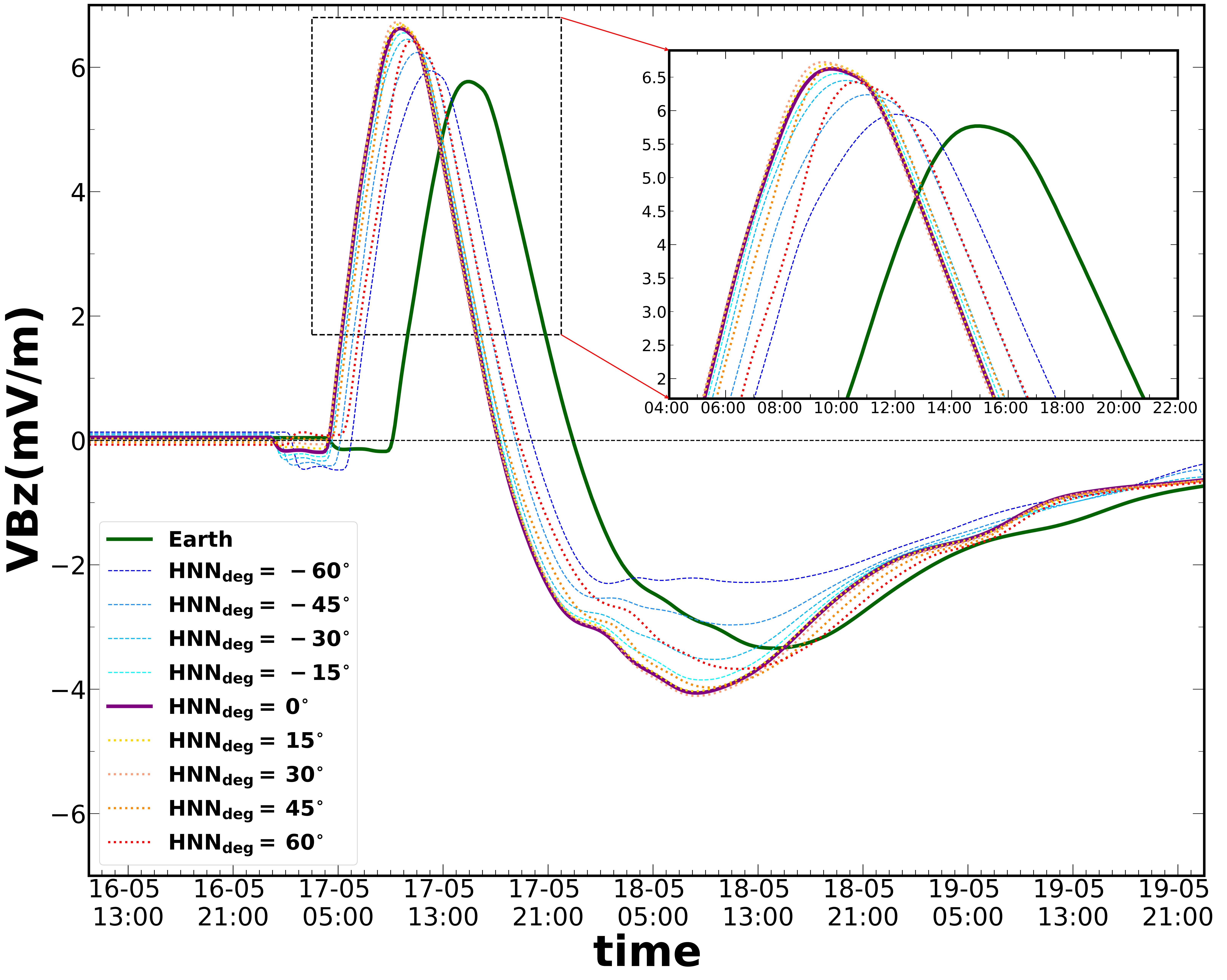}
      \caption{Same as Fig.\ref{fig5}, but for negative magnetic helicity of the flux rope (RUN2). Now, $VB_z$ is first positive and then negative; the negative values are smaller than $-3$~mV/m.
              }
         \label{fig6}
   \end{figure}

Figure~\ref{fig4} shows the same quantities as Fig.~\ref{fig3} but for the case of RUN2. The overall behaviour is similar to that of Fig.~\ref{fig3}. Due to this negative magnetic helicity, the magnetic field components inside the flux rope rotate oppositely, and therefore, the $VB_z$ is first positive and then negative. 

Figure~\ref{fig6} shows that in RUN2 $VB_z$ is first positive and then negative, and the negative values are smaller than $-3$ mV/m for an extended period. The negative values reached by $VB_z$ vary between $-2$ and $-4$~mV/m, and they are always greater than the $-5$~mV/m threshold for a geomagnetic storm. The overall behaviour is in agreement with the results of positive magnetic helicity, that is, the behaviour of $VB_z$ at HENON is very similar to that at the Earth, and the perturbation is detected first at HENON with a lead time of about 5--6 hours.
   
%-------------------------------------------------------------
   \begin{figure}[!h]
   \centering
   \includegraphics[width=0.9\textwidth]{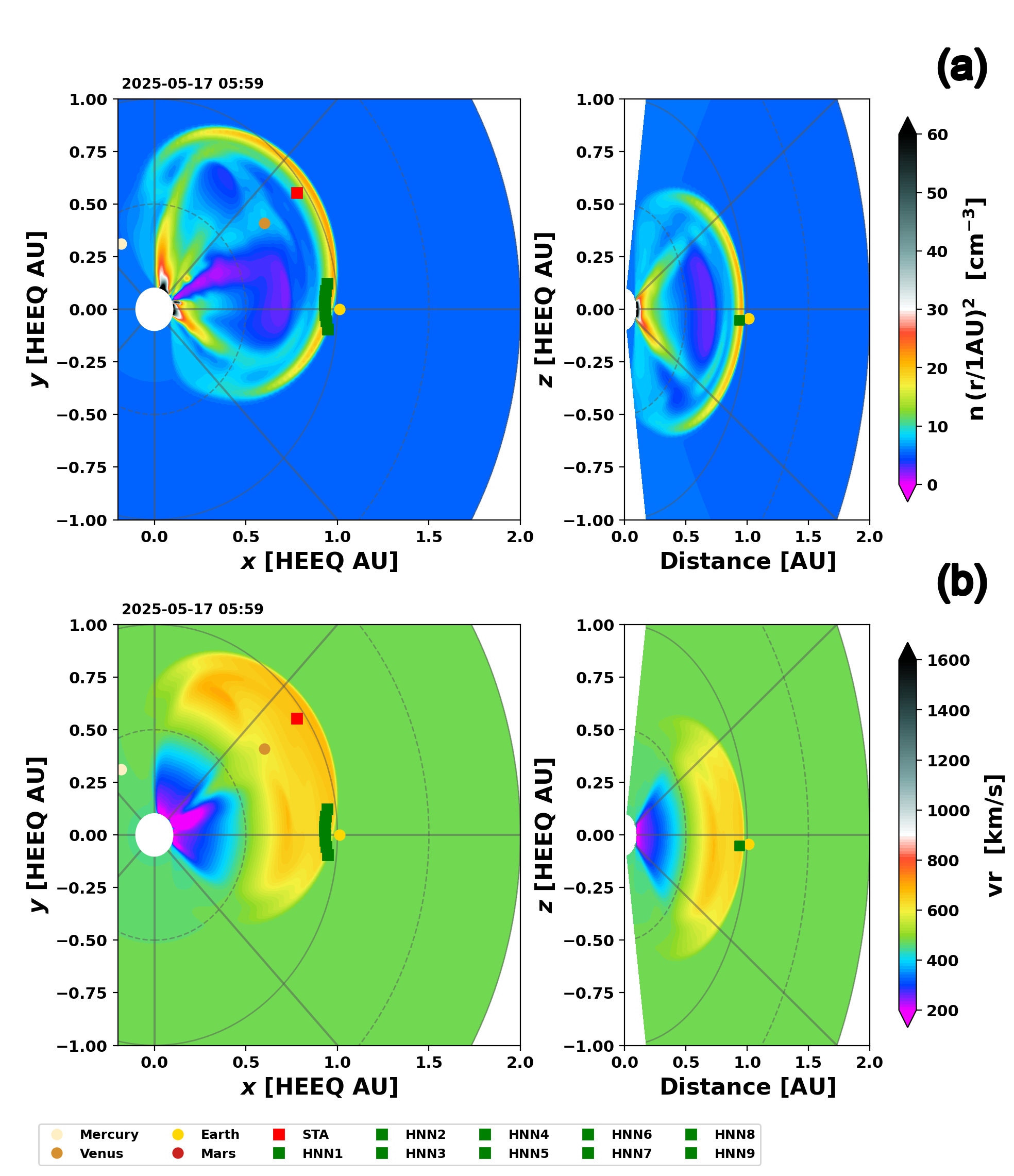}
      \caption{Simulation results RUN3, at the time when the CME is overcoming the HENON virtual spacecraft. Same format as Fig.\ref{fig2}.
              }
         \label{fig7}
   \end{figure}

We also conducted several numerical simulations for the case where the CME heliographic central longitude does not align with the Sun-Earth line.
Figure~\ref{fig7} (RUN3) shows the cross sections in the equatorial and meridional planes of the EUHFORIA simulation for density and radial component of the speed, when the CME is propagating towards $+30^\circ$. In such a case, both Earth and HENON are crossed by the CME's eastern leg. The plasma and magnetic field parameters (not shown), for the case of positive helicity, show the shock and the plasma sheath. However, the increases in plasma quantities are slightly less abrupt and have lower amplitude than in the case of propagation along the Sun-Earth line. This is expected since both the Earth and the virtual spacecraft are crossing the CME flank side, and almost {\it along} the flux rope leg (see left panel in Fig.~\ref{fig7}).

   \begin{figure}
   \centering
   \includegraphics[width=0.8\textwidth]{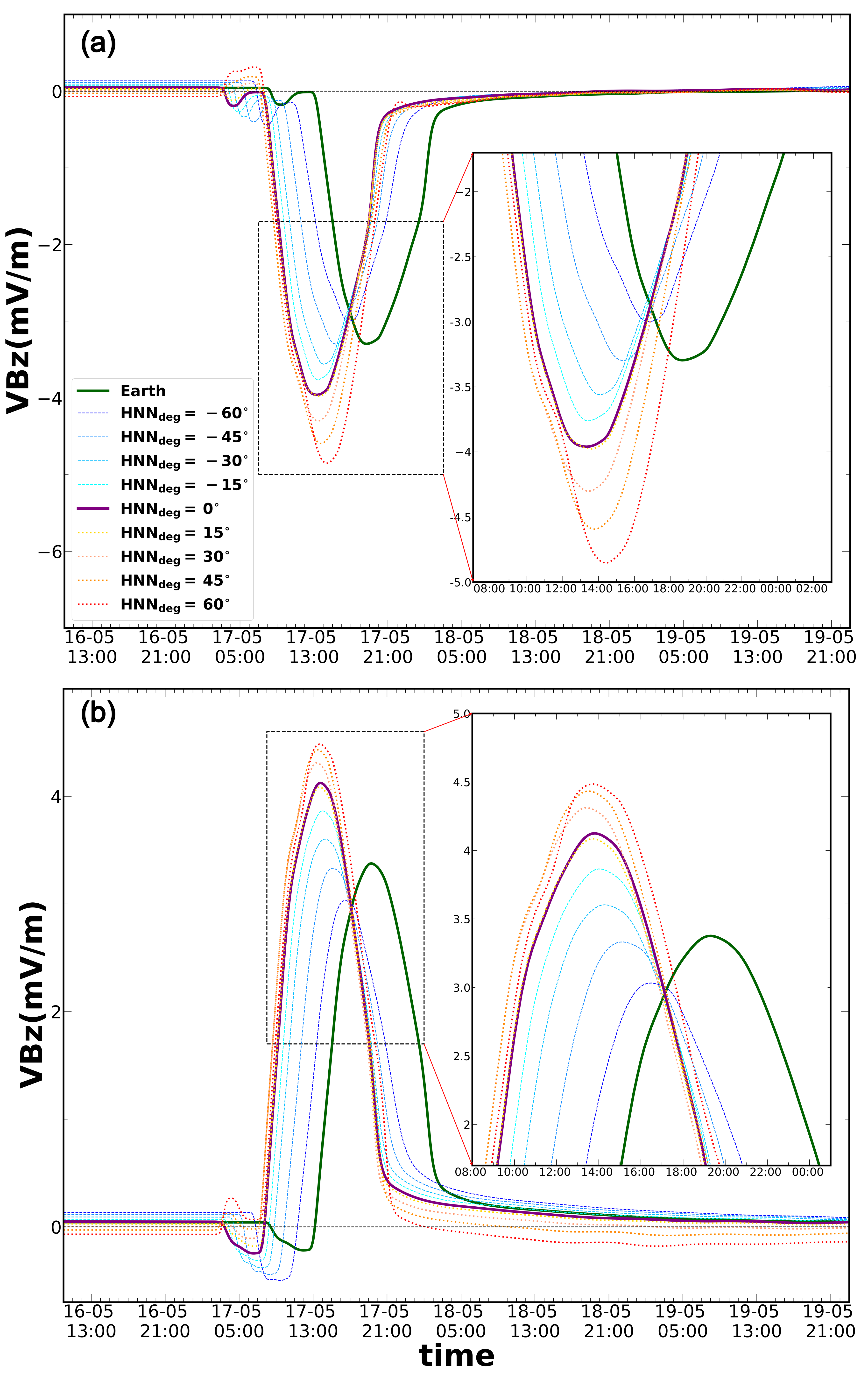}
      \caption{(a) Variation of the $VB_z$ parameter at nine virtual spacecraft along KR1 and at Earth (thick green line), for the case of a CME propagating with central apex longitude equal to $30^\circ$ W. Positive magnetic helicity is set in panel (a) (RUN3) and negative magnetic helicity is set in panel (b) (RUN4). 
             }
         \label{fig9}
   \end{figure}

   \begin{figure}[!b]
   \centering
   \includegraphics[width=0.65\textwidth]{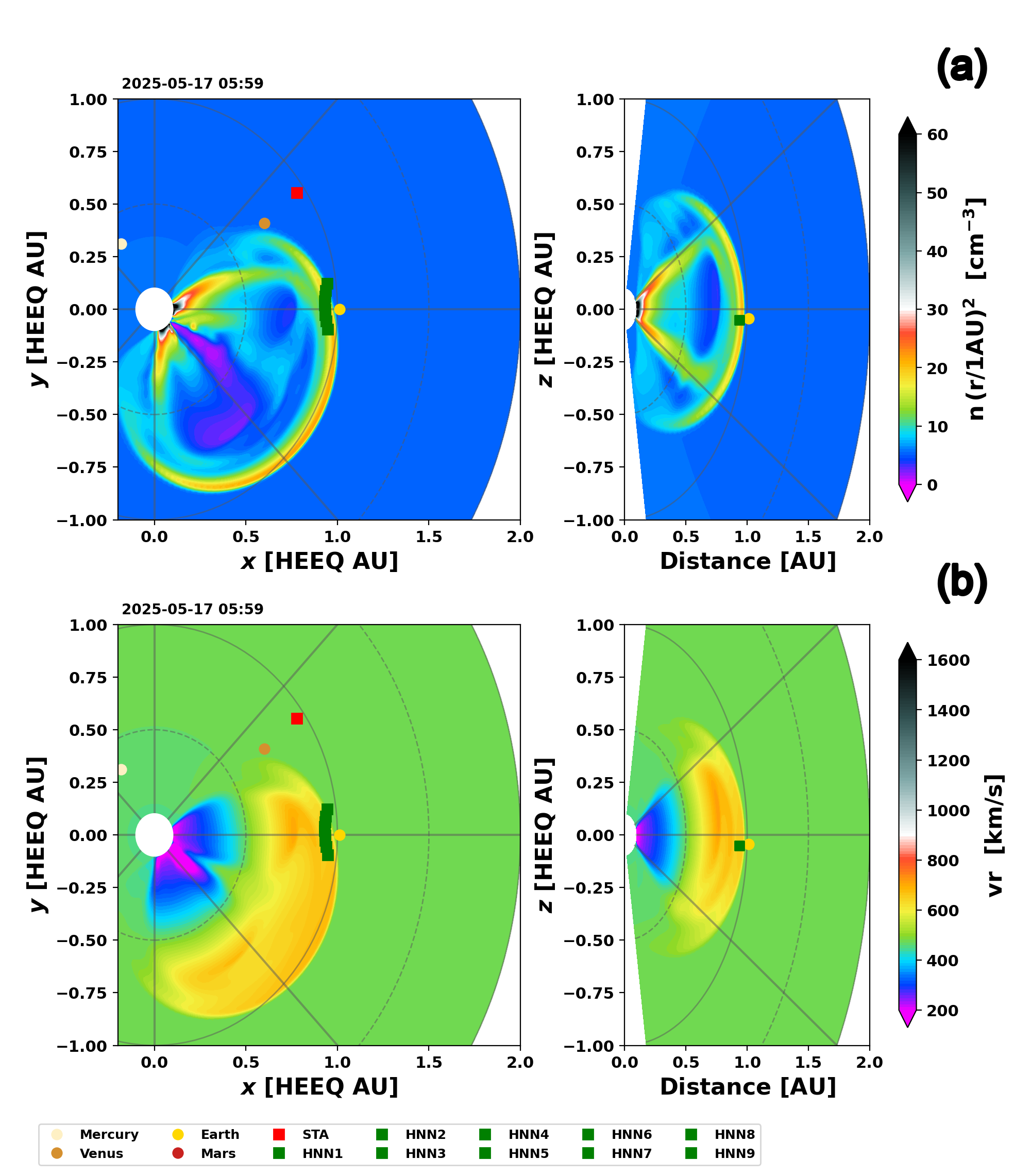}
      \caption{Simulation results for the flux rope central longitude corresponding to $30^\circ$ E, at the time when the CME is overcoming the HENON virtual spacecraft. Same format as Fig.~\ref{fig2}.
              }
         \label{fig11}
   \end{figure}

The $VB_z$ parameter is shown in Fig.~\ref{fig9} (a) for positive magnetic helicity and in Fig.~\ref{fig9} (b) for negative magnetic helicity. From Fig.~\ref{fig9} (a), we can see that the overall variations of $VB_z$ are the same at HENON and at Earth. After the initial perturbation in $VB_z$ (caused by the arrival of the shock and subsequent sheath), this quantity shows, a substantial decrease of $VB_z$ within the FRi3D magnetic flux rope is obtained. However, now the lead time varies: while for spacecraft on positive geocentric longitudes the $VB_z$ variations are almost simultaneous with (or even earlier than) those at the HENON spacecraft just in front of the Earth (solid purple line), for spacecraft located at negative geocentric longitudes the variations of $VB_z$ are progressively delayed with respect to the brown line, and the lead time with respect to the Earth is almost 2 hours when HENON is at $-60^\circ$ of geocentric longitude. This can be understood by looking at the left panel in Fig.~\ref{fig7}, which shows that the CME curved front crosses the HENON spacecraft earlier when they are on positive geocentric longitudes (even earlier than the spacecraft on the Sun-Earth line) and later when the spacecraft are on negative longitudes. We further notice that the peak values of $VB_z$ change with the HENON longitude, the perturbation being larger when HENON is closer to the CME legs.

Figure~\ref{fig9}(b) shows the corresponding results for RUN4. Now, the values of $VB_z$ are primarily positive due to the change in the magnetic helicity. The results are similar to those of RUN3, with a simultaneous variation of $VB_z$ for spacecraft located in front of the Earth and at positive geocentric longitudes, and a progressively delayed variation of $VB_z$ for those at negative geocentric longitudes.

%As for the case of positive helicity, the variations of $VB_z$ are almost simultaneous with (or even earlier than) those at the spacecraft just in front of the Earth, brown line, while for spacecraft on negative geocentric longitudes the variations of $VB_z$ are progressively delayed.
%-------------------------------------------------------------

%
We can observe that, in the case of CME central longitude equal to $30^\circ$ W, the $B_z$ measured by the various spacecraft does not change sign as a function of time (not shown), it being either entirely positive or negative: we consider this to be due to the magnetic axis of the CME leg not crossing the spacecraft.
%-------------------------------------------------------------

   \begin{figure}[!h]
   \centering
   \includegraphics[width=0.75\textwidth]{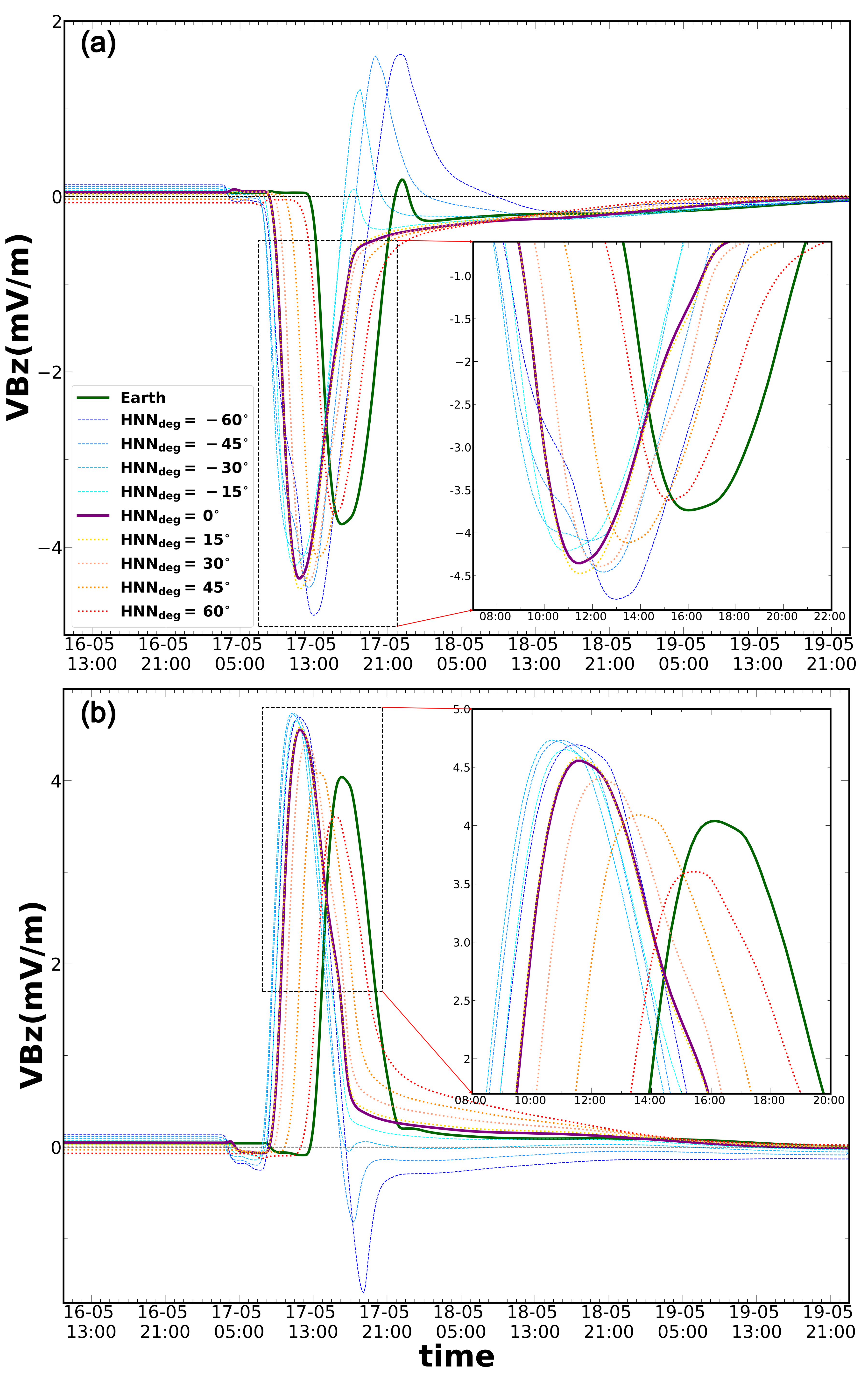}
      \caption{Variation of the $VB_z$ parameter for the case of a CME propagating with central apex longitude equal to $-30^\circ$ and positive magnetic helicity. 
      Same format as Fig.~\ref{fig9}.
              }
         \label{fig13}
   \end{figure}

We have carried out numerical experiments also for the case of the CME pointing to the east (RUN5-6), i.e., for central apex longitude equal to $30^\circ$ E: in this case the CME western leg is overcoming HENON and the Earth, and the results are similar to those obtained for RUN3-4, but with HENON spacecraft at negative geocentric longitudes seeing first the perturbation, as shown in Fig.~\ref{fig11}. The magnetic field and plasma parameters (not shown) at the HENON spacecraft on the Sun-Earth line and at the Earth are very similar, with peak values being larger at HENON.

The $VB_z$ parameter for RUN5 is shown in Fig.~\ref{fig13} (a): in this case, too, the overall behaviour is similar at HENON and at the Earth, but now the lead time is maximum ($\sim 6$ hours) for spacecraft on negative geocentric longitudes (shades of blue), while the lead time gradually decreases to about 1 hour for spacecraft on positive geocentric longitudes (shades of red). In addition, the peak values of $VB_z$ change with the position of the HENON spacecraft but remain larger than the values at the Earth, except for the spacecraft at $60^\circ$ of geocentric longitude. 

Figure~\ref{fig13} (b) shows the corresponding results for RUN6, with the values of $VB_z$ mainly being positive, due to the reversal in the sign of $B_z$ registered at the virtual spacecraft after the passage of the sheath (not shown). As for the case of positive helicity (Figure~\ref{fig13} (a)), the variations of $VB_z$ for spacecraft with negative geocentric longitudes are almost simultaneous with (or even earlier than) those at the spacecraft just in front of the Earth (solid purple line), while for spacecraft on positive geocentric longitudes the variations of $VB_z$ are progressively delayed, with the lead time reduced to about 1 hour for the spacecraft at $+60^\circ$ of geocentric longitude. 

\subsection{Predictive Capabilities of the HENON mission}

To better understand the results obtained for $VB_z$, we take the maximum of its absolute value as a function of the Alert time. We define the Alert time as the difference between the time corresponding to the maximum absolute value of $VB_z$ at each position of the HENON virtual spacecraft and the corresponding time calculated at Earth
\[
Alert \: Time = t_{HENON} - t_{Earth}.
\]
This value is of fundamental importance for the early warning and mitigation of potentially hazardous geomagnetic disturbances.

   \begin{figure}[!h]
   \centering
   \includegraphics[width=1\textwidth]{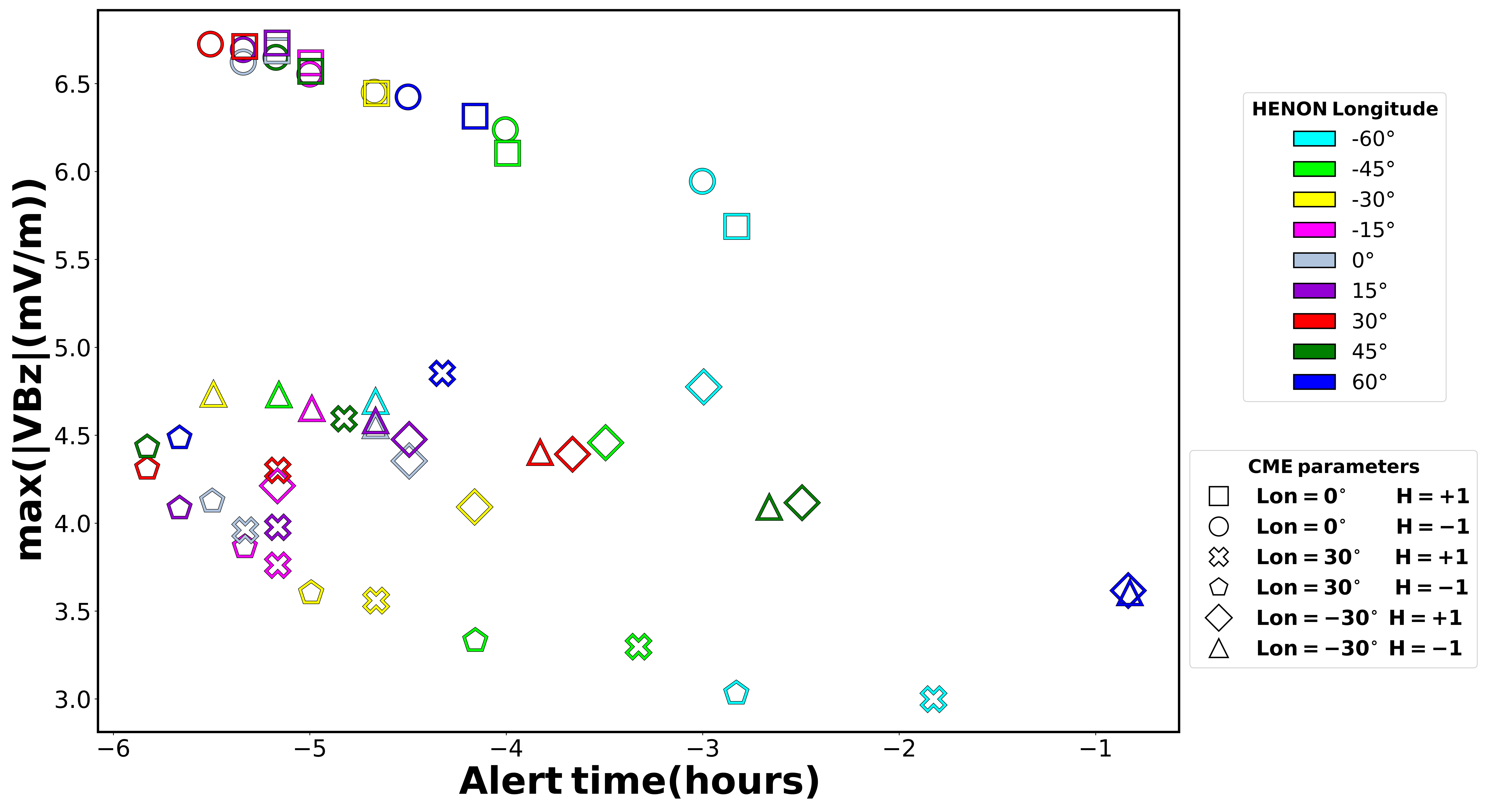}
      \caption{Alert Time at Earth for different CME events. All the CMEs have an initial speed of 750 km s$^{-1}$. The delay is measured as the difference between the time of the maximum of $|VB_z|$ recorded at the position of the HENON virtual spacecraft and the corresponding time measured at Earth. Squares and circles represent RUN1 and RUN2; crosses and pentagons represent RUN3 and RUN4; diamonds and triangles represent RUN5 and RUN6 (see legend CME parameters). The colour of each symbol indicates the spacecraft geocentric longitude (see legend).}
         \label{fig:maxVBz}
   \end{figure}

In Fig.~\ref{fig:maxVBz}, we show the maximum of the absolute value of $VB_z$ for each virtual spacecraft and for each simulation. We indicate RUN1 and RUN2 with squares and circles; RUN3 and RUN4 with crosses and pentagons; and RUN5 and RUN6 with diamonds and triangles.
For almost all configurations of CMEs and HENON positions within the KR1, the alert time is greater than 4 hours.
The largest values of max$(|VB_z|)$ are reached when the CME is launched at $0^{\circ}$ in both longitude and latitude (RUN1 and 2). The results from those runs are clustered in the top left side of the panel in Fig.~\ref{fig:maxVBz}. In such a branch, we can notice that the best Alert Times are obtained when spacecraft are located within the longitude range between $-45 ^{\circ}$ and $+60 ^{\circ}$, as the CME nose impacts these spacecraft head-on. This could be related to the interaction between the background solar wind and the CME, as the polarity of the background solar wind can influence the deformation and deflection of the CME. This may explain why, particularly away from the nose of the CME, the circles consistently show a longer Alert Time than the squares.

Several differences arise when the CME is directed westward (RUN3 and 4) or eastward (RUN5 and 6) with respect to the Sun. It is evident that, in these cases, the values of max$(|VB_z|)$ are lower than for RUN1 and 2, as these virtual spacecraft have crossed the flank of the CME. Another surprising result concerns the Alert Time values, which in most cases are also greater than the previous ones. The best values are reached in RUN4 with an alert time of about 6 hours.   

Although the max$(|VB_z|)$ values oscillate between 3.5 and 5.0 mV/m, the corresponding Alert Times differ significantly. One would expect the CME to exhibit a symmetric behaviour, giving rise to similar results in the Alert Time; however, a CME propagating westward arrives earlier than the one directed eastward. This can be related to the deflection of the CME in the interplanetary medium, as explained in \citet{wang2004deflection,Wang14,Zhuang19}. Indeed, in each run the CME speed exceeds that of the ambient solar wind, and the CME interacts with the Parker spiral field lines, which act like a “wall”. This interaction produces a slight eastward deflection, even during radial propagation. Overall, the CME is guided in the direction of the spiral winding, i.e., the Parker spiral acts as a magnetic barrier, enhancing its motion in the eastward direction. This may explain why the Alert Times are longer in the case of westward-directed CMEs than in eastward-directed ones. In each simulation, virtual spacecraft are distributed over both positive and negative longitudes. However, when a fast CME is directed westward, it reaches the spacecraft at positive longitudes earlier than in the case of an eastward-directed CME reaching spacecraft at negative longitudes. In a symmetric scenario, one would expect that, for an eastward-directed CME, the spacecraft at negative longitudes would experience similar Alert Times. The fact that this is not observed suggests that the Parker spiral influences the CME propagation, effectively deflecting it in its direction of rotation and causing an earlier arrival at spacecraft located westward.

\section{Conclusions}

In this work, we investigated the potential of the HENON mission to serve as a forward observatory, providing timely SW alerts for CME events.
HENON is a space mission designed to fly on a distant retrograde orbit around the Earth at an upstream distance of approximately 0.1~AU, allowing for advanced-time observations of SW perturbations that will impact Earth. To assess the relationship between delayed measurements at HENON and Earth, we focused on the $VB_z$ parameter, which is considered a good predictor of the geomagnetic $Dst$ index. An {\it in-situ} {\rm monitor} like HENON is crucial to measure accurate magnetic field perturbations, in particular because the geoeffectiveness of a CME strongly depends on the sign of $B_z$. 
To explore the effects of a SW event along the whole KR1 of the orbit, we performed 6 different EUHFORIA simulations using the magnetised CME model FRi3D \citep{Maharana2022}. We positioned nine virtual spacecraft in front of the Earth within the KR1, covering geocentric longitudes from $-60^\circ$ to $60^\circ$, which correspond to heliocentric longitudes from $-6.9^\circ$ to $6.9^\circ$. For the present runs, the minimum upstream distance of HENON is $0.082$~AU.  

For the cases where the central apex longitude of the CME is zero (RUN1 and 2), that is, the CME is propagating along the Sun-Earth line, the simulation results show that the overall behaviour of $VB_z$ is similar at HENON and on Earth, even when measurements are obtained at the limits (extremal longitudes) of KR1. This confirms the efficacy of the HENON mission for predictive purposes. With a positive helicity of the magnetic flux rope, the local magnetic field $B_z$ is first negative and then positive. With the simulation parameters adopted, the negative values of $VB_z$ reach $-5$~mV/m for more than three hours, indicating that a geomagnetic storm is likely to occur. Thereafter, a prolonged period of positive $VB_z$ is obtained. According to the simulation results, a CME with positive helicity (RUN1) values appears to be more hazardous than one with negative helicity values (RUN2).
As a consequence of the mission design, the variations of $VB_z$ are detected at HENON with about  3.5 to 5.5 hours of lead time for a CME propagating at more than 600 km/s along the radial direction. This represents a substantial improvement with respect to the observations carried out by spacecraft at L1. It is also found that the peak values of $VB_z$ (positive and negative) are higher at HENON, compared to the Earth; this is understood as the transverse components of the interplanetary magnetic field scale with the heliocentric distance as $\sim 1/r$. Therefore, when using the threshold for the onset of geomagnetic storms, typically $VB_z<-5$ mV/m, one should take into account that the magnetic field transverse components at HENON are larger than those at Earth by a factor that depends on the heliocentric distance of HENON ($r_{\rm HENON}$). In other words, to compare $VB_z$ with the typical thresholds at Earth, the magnetic field should be multiplied by the factor $r_{\rm HENON}/r_{\rm Earth} < 1$. We point out, however, that other scalings for the magnetic field intensity in the magnetic flux rope of CMEs have been found \citep[e.g.,][]{Davies21}, a possibility which should be taken into account. The radial dependence of the magnetic field magnitude has also been extensively investigated and reported to scale roughly as $r^{-1.6}$ in the inner heliosphere up to 1 AU \citep{maruca2022radial}.

The lead time is approximately 5.5 hours when HENON is on the Sun-Earth line; it can be seen that this lead time decreases to about 3.5 hours when HENON is at $\pm 6.9^\circ$ of heliocentric longitudes, corresponding to $\pm 60^\circ$ of geocentric longitudes. We consider that this decrease in the advance time is due to the curved shape of the trajectory as well as to the curved CME front. These variations should also be taken into account when using the forthcoming HENON data for geomagnetic storm predictions. 
For the cases when the CME apex central longitude is not pointing directly to the Earth (RUN3 and 6), but is at $\pm30^\circ$ as in Figs. \ref{fig7}--\ref{fig11}, the shape of the perturbed $VB_z$ is similar at HENON and at the Earth, but now the lead time varies: it is maximum, e.g., 6 hours or more, when HENON is on the longitudes closest to the CME main direction of propagation and up to the spacecraft on the Sun-Earth line; then the lead time gradually decreases, reaching a minimum of about 1--2 hours for the case when HENON is on the edges of KR1, farthest from the CME trajectory. Therefore, in addition to knowing the HENON position on the orbit, it is crucial to understand the propagation direction of the CME. 

Finally, we constructed a plot of the maximum value of $|VB_z|$ as a function of the Alert Time. The Alert Time is a relevant parameter derived in this work, as it provides an estimate of the signal delay between Earth and the position of the virtual spacecraft. Furthermore, we found that changing the CME's longitude affects the signal detected by the virtual spacecraft: contrary to expectations, minor differences emerge depending on the CME’s direction.
In fact, in the case of a westward-propagating CME, the virtual spacecraft located at positive longitudes exhibits longer Alert Times compared to the case of an eastward-directed CME. In a symmetric scenario, we would expect similar behaviour for spacecraft located at negative longitudes, but this does not appear to occur. This asymmetry can be explained by the deflection in the CME propagation described by \citet{wang2004deflection,Wang14,Zhuang19}, who showed that a CME interacting with the interplanetary magnetic field will be deflected toward east when it propagates faster than the background solar wind, and toward west when it propagates slower than the background solar wind, due to the pileup of solar wind plasma ahead of or behind the CME. 
In the present simulations the CME speed is of the order of 700 km/s during the whole propagation outwards, while the background solar wind speed is below 500 km/s. Because the CME propagates faster than the ambient solar wind, an eastward deflection is observed. This is evident in Fig.~\ref{fig2}, which is slightly asymmetric and oriented eastward.

In this work, we present the simulation results for several cases. Clearly, different parameters like the flux rope tilt, the central apex longitude and latitude could be adopted, also keeping in mind that CMEs frequently are not centred on the heliographic equator \citep[e.g.,][]{Zimbardo23}. When the CME is close to the heliographic equator and the central apex longitude is between $0^\circ$ and $\pm 30^\circ$, we can expect that the results for $VB_z$ are intermediate between those shown here. 
Another issue to consider is that the initial CME magnetic flux rope adopted in the simulations has a well-structured magnetic field configuration, which is reflected in the regularity and smoothness of the numerical results we have shown. However, real CMEs have a more complex and irregular magnetic structure, and even are deformed by their interaction with the background solar wind \citep{2024Mayank}, which could lead to larger differences in the $VB_z$ profiles at the Earth and HENON, compared to those reported here. 
It should also be kept in mind that numerical simulations have to make a number of simplifying assumptions. Therefore, the simulation results are important to develop the measurement  scenario of the HENON project, but some degree of uncertainty regarding the above results should be taken into account. In addition, global heliospheric simulations cannot account for the small-scale interactions of CMEs with the turbulence of the interplanetary medium, which can lead to deformations and diffusion of structures on small scales \citep{Sangalli25}. 

We now note the following: a future development of the HENON space mission for continuous SW monitoring requires flying four spacecraft on the DRO, so that at least one spacecraft will always be in the KR1 orbit. This means that when one spacecraft is close to the limits of KR1 and on the opposite side of a CME propagating at, say, $30^\circ$ W, there will be another spacecraft in KR2 (see Fig.~\ref{fig1}) that will be in a favourable position to observe the passing CME.  For instance, referring to Fig.~\ref{fig9}, a spacecraft in KR2 at geocentric longitudes between $60^\circ$ and $90^\circ$, close to the CME, would observe a $VB_z$ with a timing and intensity similar to those observed by a nearby spacecraft in KR1 at $45^\circ$ or $60^\circ$ of geocentric longitude. Such observations would have a lead time of $\sim 6$ h, and would be very useful for geomagnetic storm predictions. Therefore, even a spacecraft in KR2 can play a significant role in SW, although the presence of small scale structures, as discussed above, should be considered.  
Further simulations exploring different CME speeds, heliocentric longitudes and flux ropes tilts are planned and will be included in a dedicated database and described in a forthcoming paper.

These findings highlight that a robust space weather prediction infrastructure must incorporate, beside upstream \it in-situ \rm monitors like HENON, continuous monitoring of solar eruptive activity \citep{Vourlidas19}. This is essential for accurately determining the relative positions of HENON and any propagating CME, as our analysis demonstrates that the lead times, peak values of $VB_z$, and the feasibility of utilising data from a spacecraft in KR2 depend critically on this relative positioning.
We can conclude that the propagation direction of a CME is of fundamental importance in determining the potential severity of a geomagnetic storm. As shown by the simulations carried out, it is possible to issue a significant alert with a lead time of up to 6 hours when the CME is directed frontally toward the Earth. An alert of a few hours is also possible even when the CME is not precisely aligned with the Sun–Earth line. This represents an excellent result for geomagnetic storm forecasting, as it allows us to predict well in advance whether a specific event may pose a risk to space weather–related activities. 

The increasing demand for reliable space weather forecasts necessitates a concerted effort from the heliophysics community to develop synergistic approaches that integrate both in-situ missions and remote observatories \citep{long2023eruption}. 
Current missions, such as Parker Solar Probe and Solar Orbiter, have significantly improved our knowledge of solar dynamics through {\it in-situ} observations near the Sun. Upcoming missions at L4 and L5, such as Vigil, will help to improve early detection of Earth-directed solar events. Polarized images from the new PUNCH mission can discern the direction of propagation of CMEs at their early onset, providing timely alerts for the whole network of remote and {\it in-situ} detectors such as HENON. 
Notably, the imminent and unprecedented observations provided by the PUNCH mission perfectly complement the in situ measurements from HENON, thereby enhancing our collective predictive capabilities.

\section*{Funding}

This work is the outcome of the preparatory phases for the launch of the HENON mission by the Italian Space Agency (ASI) and was partially funded through the Argotec contracts, numbers ARG-IT-CON-P-HEN-220002 and ARG-IT-CON-P-HEN-250003. HENON is part of the ASI program Alcor and is being developed under the European Space Agency's General Support Technology Programme (ESA-GSTP) through the support of the national delegations of Italy (ASI), the UK, Finland, and the Czech Republic.
GZ, S. Perri, and GP acknowledge partial support by the Italian PRIN 2022, project 2022294WNB entitled "Heliospheric shocks and space weather: from multispacecraft observations to numerical modelling” (CUP H53D23000900006), funded by Next Generation EU, fondo del Piano Nazionale di Ripresa e Resilienza (PNRR) Missione 4 “Istruzione e Ricerca” - Componente C2 Investimento 1.1, ‘Fondo per il Programma Nazionale di Ricerca e Progetti di Rilevante Interesse Nazionale (PRIN). 
S. Perri, GN, FC and GZ acknowledge the project ‘Data-based predictions of solar energetic particle arrival to the Earth: ensuring space data and technology integrity from hazardous solar activity events’ (CUP H53D23011020001) ‘Finanziato dall’Unione europea – Next Generation EU’ PIANO NAZIONALE
DI RIPRESA E RESILIENZA (PNRR) Missione 4 “Istruzione e Ricerca” - Componente C2 Investimento 1.1, ‘Fondo per il Programma Nazionale di Ricerca e
Progetti di Rilevante Interesse Nazionale (PRIN)’ Settore PE09. %AG, FL, S. Perri, GP, GN, SS, FM and GZ were further supported by the Space It Up project funded by the Italian Space Agency, ASI, and the Ministry of University and Research, MUR, under contract n. 2024-5-E.0 - CUP n. I53D24000060005.
FP acknowledges support by NASA/SWRI PUNCH subcontract N99054DS at the University of Delaware, and a Plan for NASA EPSCoR Research Infrastructure Development (RID) in Delaware (NASA award 80NSSC22M0039).
SP is funded by the European Union. Views and opinions expressed are, however, those of the author(s) only and do not necessarily reflect those of the European Union or ERCEA. Neither the European Union nor the granting authority can be held responsible for them. This project (Open SESAME) has received funding under the Horizon Europe programme (ERC-AdG agreement No 101141362). These results were also obtained in the framework of the projects C16/24/010 (C1 project Internal Funds KU Leuven), G0B5823N and G002523N (WEAVE) (FWO-Vlaanderen), and 4000145223 (SIDC Data Exploitation (SIDEX2), ESA Prodex).\\ 
SS aknowledges support from the Space It Up project funded by the Italian Space Agency, ASI, and the Ministry of University and Research, MUR, under Contract Grant Nos. 2024-5-E.0-CUP and I53D24000060005.

\section*{Conflict of interest}
The authors declare no conflict of interest.
%%    This version assumes use of bibtex with the jswsc.bib file being present
%%    If your bib file has a different name you need to change the following line
\section*{Data availability statement}

The authors declare that no data are associated with this article.

\bibliography{jswsc}

@ARTICLE{Cyr00,
       author = {{St. Cyr}, O.~C. and {Mesarch}, M.~A. and {Maldonado}, H.~M. and {Folta}, D.~C. and {Harper}, A.~D. and {Davila}, J.~M. and {Fisher}, R.~R.},
        title = "{Space Weather Diamond: a four spacecraft monitoring system}",
      journal = {Journal of Atmospheric and Solar-Terrestrial Physics},
         year = 2000,
        month = sep,
       volume = {62},
       number = {14},
        pages = {1251-1255},
          doi = {10.1016/S1364-6826(00)00069-9},
       adsurl = {https://ui.adsabs.harvard.edu/abs/2000JASTP..62.1251S}
}

@ARTICLE{Eastwood24,
       author = {{Eastwood}, J.~P. and {Brown}, P. and {Magnes}, W. and {Carr}, C.~M. and {Agu}, M. and {Baughen}, R. and {Berghofer}, G. and {Hodgkins}, J. and {Jernej}, I. and {M{\"o}stl}, C. and {Oddy}, T. and {Strickland}, A. and {Vitkova}, A.},
        title = "{The Vigil Magnetometer for Operational Space Weather Services From the Sun-Earth L5 Point}",
      journal = {Space Weather},
         year = 2024,
        month = jun,
       volume = {22},
       number = {6},
          eid = {e2024SW003867},
        pages = {e2024SW003867},
          doi = {10.1029/2024SW003867},
       adsurl = {https://ui.adsabs.harvard.edu/abs/2024SpWea..2203867E}
}

@ARTICLE{Palmerio25,
       author = {{Palmerio}, Erika},
        title = "{Monitoring the Solar Wind Before It Reaches L1}",
      journal = {Space Weather},
         year = 2025,
        month = nov,
       volume = {23},
       number = {11},
          eid = {e2025SW004452},
        pages = {e2025SW004452},
          doi = {10.1029/2025SW004452},
archivePrefix = {arXiv},
       eprint = {2511.08463},
 primaryClass = {physics.space-ph},
       adsurl = {https://ui.adsabs.harvard.edu/abs/2025SpWea..2304452P}
}

@ARTICLE{Lugaz25,
       author = {{Lugaz}, No{\'e} and {Al-Haddad}, Nada and {Zhuang}, Bin and {M{\"o}stl}, Christian and {Davies}, Emma E. and {Farrugia}, Charles J. and {Banu}, Sahanaj Aktar and {Weiler}, Eva and {Galvin}, Antoinette B.},
        title = "{The Need for a Sub-L1 Space Weather Research Mission: Current Knowledge Gaps on Coronal Mass Ejections}",
      journal = {Space Weather},
         year = 2025,
        month = feb,
       volume = {23},
       number = {2},
        pages = {2024SW004189},
          doi = {10.1029/2024SW004189},
       adsurl = {https://ui.adsabs.harvard.edu/abs/2025SpWea..2304189L}
}

@ARTICLE{Lugaz24b,
       author = {{Lugaz}, No{\'e} and {Lee}, Christina O. and {Al-Haddad}, Nada and {Lillis}, Robert J. and {Jian}, Lan K. and {Curtis}, David W. and {Galvin}, Antoinette B. and {Whittlesey}, Phyllis L. and {Rahmati}, Ali and {Zesta}, Eftyhia and {Moldwin}, Mark and {Summerlin}, Errol J. and {Larson}, Davin E. and {Courtade}, Sasha and {French}, Richard and {Hunter}, Richard and {Covitti}, Federico and {Cosgrove}, Daniel and {Prall}, J.~D. and {Allen}, Robert C. and {Zhuang}, Bin and {Winslow}, R{\'e}ka M. and {Scolini}, Camilla and {Lynch}, Benjamin J. and {Filwett}, Rachael J. and {Palmerio}, Erika and {Farrugia}, Charles J. and {Smith}, Charles W. and {M{\"o}stl}, Christian and {Weiler}, Eva and {Janvier}, Miho and {Regnault}, Florian and {Livi}, Roberto and {Nieves-Chinchilla}, Teresa},
        title = "{The Need for Near-Earth Multi-Spacecraft Heliospheric Measurements and an Explorer Mission to Investigate Interplanetary Structures and Transients in the Near-Earth Heliosphere}",
      journal = {\ssr},
         year = 2024,
        month = oct,
       volume = {220},
       number = {7},
          eid = {73},
        pages = {73},
          doi = {10.1007/s11214-024-01108-8},
       adsurl = {https://ui.adsabs.harvard.edu/abs/2024SSRv..220...73L}
}

@ARTICLE{Lugaz24a,
       author = {{Lugaz}, No{\'e} and {Zhuang}, Bin and {Scolini}, Camilla and {Al-Haddad}, Nada and {Farrugia}, Charles J. and {Winslow}, R{\'e}ka M. and {Regnault}, Florian and {M{\"o}stl}, Christian and {Davies}, Emma E. and {Galvin}, Antoinette B.},
       title = "{The Width of Magnetic Ejecta Measured near 1 au: Lessons from STEREO-A Measurements in 2021--2022}",
       journal = {\apj},
       year = 2024,
       month = feb,
       volume = {962},
       number = {2},
       eid = {193},
       pages = {193},
       doi = {10.3847/1538-4357/ad17b9},
       archivePrefix = {arXiv},
       eprint = {2312.03942},
       primaryClass = {astro-ph.SR},
       adsurl = {https://ui.adsabs.harvard.edu/abs/2024ApJ...962..193L}
}

@ARTICLE{Lugaz18,
       author = {{Lugaz}, No{\'e} and {Farrugia}, Charles J. and {Winslow}, Reka M. and {Al-Haddad}, Nada and {Galvin}, Antoinette B. and {Nieves-Chinchilla}, Teresa and {Lee}, Christina O. and {Janvier}, Miho},
        title = "{On the Spatial Coherence of Magnetic Ejecta: Measurements of Coronal Mass Ejections by Multiple Spacecraft Longitudinally Separated by 0.01 au}",
      journal = {The Astrophysical Journal Letters},
         year = 2018,
        month = sep,
       volume = {864},
       number = {1},
          eid = {L7},
        pages = {L7},
          doi = {10.3847/2041-8213/aad9f4},
archivePrefix = {arXiv},
       eprint = {1812.00911},
 primaryClass = {physics.space-ph},
       adsurl = {https://ui.adsabs.harvard.edu/abs/2018ApJ...864L...7L}
}

@ARTICLE{Scolini24,
       author = {{Scolini}, Camilla and {Lugaz}, No{\'e} and {Winslow}, R{\'e}ka M. and {Farrugia}, Charles J. and {Magyar}, Norbert and {Bacchini}, Fabio},
        title = "{On the Role of Alfv{\'e}nic Fluctuations as Mediators of Coherence within Interplanetary Coronal Mass Ejections: Investigation of Multi-spacecraft Measurements at 1 au}",
      journal = {\apj},
         year = 2024,
        month = jan,
       volume = {961},
       number = {1},
          eid = {135},
        pages = {135},
          doi = {10.3847/1538-4357/ad0ed1},
archivePrefix = {arXiv},
       eprint = {2312.04480},
 primaryClass = {astro-ph.SR},
       adsurl = {https://ui.adsabs.harvard.edu/abs/2024ApJ...961..135S}
}

@ARTICLE{Laker24,
       author = {{Laker}, R. and {Horbury}, T.~S. and {O'Brien}, H. and {Fauchon-Jones}, E.~J. and {Angelini}, V. and {Fargette}, N. and {Amerstorfer}, T. and {Bauer}, M. and {M{\"o}stl}, C. and {Davies}, E.~E. and {Davies}, J.~A. and {Harrison}, R. and {Barnes}, D. and {Dumbovi{\'c}}, M.},
        title = "{Using Solar Orbiter as an Upstream Solar Wind Monitor for Real Time Space Weather Predictions}",
      journal = {Space Weather},
         year = 2024,
        month = feb,
       volume = {22},
       number = {2},
          eid = {e2023SW003628},
        pages = {e2023SW003628},
          doi = {10.1029/2023SW003628},
archivePrefix = {arXiv},
       eprint = {2307.01083},
 primaryClass = {physics.space-ph},
       adsurl = {https://ui.adsabs.harvard.edu/abs/2024SpWea..2203628L}
}

@ARTICLE{Weiler25,
       author = {{Weiler}, E. and {M{\"o}stl}, C. and {Davies}, E.~E. and {Veronig}, A.~M. and {Amerstorfer}, U.~V. and {Amerstorfer}, T. and {Le Lou{\"e}dec}, J. and {Bauer}, M. and {Lugaz}, N. and {Haberle}, V. and {R{\"u}disser}, H.~T. and {Majumdar}, S. and {Reiss}, M.},
        title = "{First Observations of a Geomagnetic Superstorm With a Sub-L1 Monitor}",
      journal = {Space Weather},
         year = 2025,
        month = mar,
       volume = {23},
       number = {3},
        pages = {2024SW004260},
          doi = {10.1029/2024SW004260},
archivePrefix = {arXiv},
       eprint = {2411.12490},
 primaryClass = {physics.space-ph},
       adsurl = {https://ui.adsabs.harvard.edu/abs/2025SpWea..2304260W}
}

@ARTICLE{Borovsky14,
       author = {{Borovsky}, Joseph E. and {Birn}, Joachim},
        title = "{The solar wind electric field does not control the dayside reconnection rate}",
      journal = {Journal of Geophysical Research (Space Physics)},
         year = 2014,
        month = feb,
       volume = {119},
       number = {2},
        pages = {751-760},
          doi = {10.1002/2013JA019193},
       adsurl = {https://ui.adsabs.harvard.edu/abs/2014JGRA..119..751B}
}

@ARTICLE{Temmer15,
       author = {{Temmer}, M. and {Nitta}, N.~V.},
        title = "{Interplanetary Propagation Behavior of the Fast Coronal Mass Ejection on 23 July 2012}",
      journal = {\solphys},
         year = 2015,
        month = mar,
       volume = {290},
       number = {3},
        pages = {919-932},
          doi = {10.1007/s11207-014-0642-3},
archivePrefix = {arXiv},
       eprint = {1411.6559},
 primaryClass = {astro-ph.SR},
       adsurl = {https://ui.adsabs.harvard.edu/abs/2015SoPh..290..919T}
}

@ARTICLE{Liu14,
       author = {{Liu}, Ying D. and {Luhmann}, Janet G. and {Kajdi{\v{c}}}, Primo{\v{z}} and {Kilpua}, Emilia K.~J. and {Lugaz}, No{\'e} and {Nitta}, Nariaki V. and {M{\"o}stl}, Christian and {Lavraud}, Benoit and {Bale}, Stuart D. and {Farrugia}, Charles J. and {Galvin}, Antoinette B.},
        title = "{Observations of an extreme storm in interplanetary space caused by successive coronal mass ejections}",
      journal = {Nature Communications},
         year = 2014,
        month = mar,
       volume = {5},
          eid = {3481},
        pages = {3481},
          doi = {10.1038/ncomms4481},
archivePrefix = {arXiv},
       eprint = {1405.6088},
 primaryClass = {physics.space-ph},
       adsurl = {https://ui.adsabs.harvard.edu/abs/2014NatCo...5.3481L}
}

@ARTICLE{Davies21,
       author = {{Davies}, Emma E. and {Forsyth}, Robert J. and {Winslow}, R{\'e}ka M. and {M{\"o}stl}, Christian and {Lugaz}, No{\'e}},
        title = "{A Catalog of Interplanetary Coronal Mass Ejections Observed by Juno between 1 and 5.4 au}",
      journal = {\apj},
         year = 2021,
        month = dec,
       volume = {923},
       number = {2},
          eid = {136},
        pages = {136},
          doi = {10.3847/1538-4357/ac2ccb},
archivePrefix = {arXiv},
       eprint = {2111.11336},
 primaryClass = {physics.space-ph},
       adsurl = {https://ui.adsabs.harvard.edu/abs/2021ApJ...923..136D}
}

@ARTICLE{Shen14,
       author = {{Shen}, Chenglong and {Wang}, Yuming and {Pan}, Zonghao and {Miao}, Bin and {Ye}, Pinzhong and {Wang}, S.},
        title = "{Full-halo coronal mass ejections: Arrival at the Earth}",
      journal = {Journal of Geophysical Research (Space Physics)},
         year = 2014,
        month = jul,
       volume = {119},
       number = {7},
        pages = {5107-5116},
          doi = {10.1002/2014JA020001},
archivePrefix = {arXiv},
       eprint = {1406.4589},
 primaryClass = {astro-ph.SR},
       adsurl = {https://ui.adsabs.harvard.edu/abs/2014JGRA..119.5107S}
}

@ARTICLE{Wang14,
       author = {{Wang}, Yuming and {Wang}, Boyi and {Shen}, Chenglong and {Shen}, Fang and {Lugaz}, No{\'e}},
        title = "{Deflected propagation of a coronal mass ejection from the corona to interplanetary space}",
      journal = {Journal of Geophysical Research (Space Physics)},
         year = 2014,
        month = jul,
       volume = {119},
       number = {7},
        pages = {5117-5132},
          doi = {10.1002/2013JA019537},
archivePrefix = {arXiv},
       eprint = {1406.4684},
 primaryClass = {physics.space-ph},
       adsurl = {https://ui.adsabs.harvard.edu/abs/2014JGRA..119.5117W}
}

@ARTICLE{Zhuang19,
       author = {{Zhuang}, Bin and {Wang}, Yuming and {Hu}, Youqiu and {Shen}, Chenglong and {Liu}, Rui and {Gou}, Tingyu and {Zhang}, Quanhao and {Li}, Xiaolei},
        title = "{Numerical Simulations on the Deflection of Coronal Mass Ejections in the Interplanetary Space}",
      journal = {\apj},
         year = 2019,
        month = may,
       volume = {876},
       number = {1},
          eid = {73},
        pages = {73},
          doi = {10.3847/1538-4357/ab139e},
       adsurl = {https://ui.adsabs.harvard.edu/abs/2019ApJ...876...73Z}
}

@article{brunocarbone2005review,
  title={The solar wind as a turbulence laboratory},
  author={Bruno, Roberto and Carbone, Vincenzo},
  journal={Living Reviews in Solar Physics},
  volume={10},
  number={1},
  pages={2},
  year={2013},
  doi={https://doi.org/10.12942/lrsp-2013-2},
  publisher={Springer}
}

@article{matthaeus2015intermittency,
author = {Matthaeus, W. H.  and Wan, Minping  and Servidio, S.  and Greco, A.  and Osman, K. T.  and Oughton, S.  and Dmitruk, P. },
title = {Intermittency, nonlinear dynamics and dissipation in the solar wind and astrophysical plasmas},
journal = {Philosophical Transactions of the Royal Society A: Mathematical, Physical and Engineering Sciences},
volume = {373},
number = {2041},
pages = {20140154},
year = {2015},
doi = {10.1098/rsta.2014.0154},
URL = {https://royalsocietypublishing.org/doi/abs/10.1098/rsta.2014.0154}
}

@article{long2023eruption,
doi = {10.3847/1538-4357/acefd5},
url = {https://dx.doi.org/10.3847/1538-4357/acefd5},
year = {2023},
month = {9},
publisher = {The American Astronomical Society},
volume = {955},
number = {2},
pages = {152},
author = {David M. Long and Lucie M. Green and Francesco Pecora and David H. Brooks and Hanna Strecker and David Orozco-Suárez and Laura A. Hayes and Emma E. Davies and Ute V. Amerstorfer and Marilena Mierla and David Lario and David Berghmans and Andrei N. Zhukov and Hannah T. Rüdisser},
title = {The Eruption of a Magnetic Flux Rope Observed by Solar Orbiter and Parker Solar Probe},
journal = {The Astrophysical Journal}
}

@inbook{gosling1990coronal,
author = {Gosling, J. T.},
publisher = {American Geophysical Union (AGU)},
isbn = {9781118663868},
title = {Coronal Mass Ejections and Magnetic Flux Ropes in Interplanetary Space},
booktitle = {Physics of Magnetic Flux Ropes},
chapter = {},
pages = {343-364},
doi = {https://doi.org/10.1029/GM058p0343},
year = {1990}
}

@ARTICLE{maruca2023trans,
       author = {{Maruca}, Bennett A. and {Qudsi}, Ramiz A. and {Alterman}, B.~L. and {Walsh}, Brian M. and {Korreck}, Kelly E. and {Verscharen}, Daniel and {Bandyopadhyay}, Riddhi and {Chhiber}, Rohit and {Chasapis}, Alexandros and {Parashar}, Tulasi N. and {Matthaeus}, William H. and {Goldstein}, Melvyn L.},
        title = "{The Trans-Heliospheric Survey. Radial trends in plasma parameters across the heliosphere}",
      journal = {{A\&A}},
         year = 2023,
        month = jul,
       volume = {675},
        pages = {A196},
          doi = {10.1051/0004-6361/202345951}
}

@ARTICLE{Gonzalez89,
       author = {{Gonzalez}, Walter D. and {Tsurutani}, Bruce T. and {Gonzalez}, Alicia L.~C. and {Smith}, Edward J. and {Tang}, Frances and {Akasofu}, Syun-I.},
        title = "{Solar wind-magnetosphere coupling during intense magnetic storms (1978-1979)}",
      journal = {J. Geophys. Res.},
         year = 1989,
        month = jul,
       volume = {94},
       number = {A7},
        pages = {8835-8851},
          doi = {10.1029/JA094iA07p08835},
       adsurl = {https://ui.adsabs.harvard.edu/abs/1989JGR....94.8835G}
}

@ARTICLE{Gonzalez94,
       author = {{Gonzalez}, W.~D. and {Joselyn}, J.~A. and {Kamide}, Y. and {Kroehl}, H.~W. and {Rostoker}, G. and {Tsurutani}, B.~T. and {Vasyliunas}, V.~M.},
        title = "{What is a geomagnetic storm?}",
      journal = {J. Geophys. Res.},
         year = 1994,
        month = apr,
       volume = {99},
       number = {A4},
        pages = {5771-5792},
          doi = {10.1029/93JA02867},
       adsurl = {https://ui.adsabs.harvard.edu/abs/1994JGR....99.5771G}
}

@ARTICLE{Gopalswamy22,
       author = {{Gopalswamy}, N. and {Yashiro}, S. and {Akiyama}, S. and {Xie}, H. and {M{\"a}kel{\"a}}, P. and {Fok}, M. -C. and {Ferradas}, C.~P.},
        title = "{What Is Unusual About the Third Largest Geomagnetic Storm of Solar Cycle 24?}",
      journal = {Journal of Geophysical Research (Space Physics)},
         year = 2022,
        month = aug,
       volume = {127},
       number = {8},
          eid = {e30404},
        pages = {e30404},
          doi = {10.1029/2022JA030404},
archivePrefix = {arXiv},
       eprint = {2207.11630},
 primaryClass = {astro-ph.SR},
       adsurl = {https://ui.adsabs.harvard.edu/abs/2022JGRA..12730404G}
}

@ARTICLE{Henon69,
       author = {{Henon}, M.},
        title = "{Numerical exploration of the restricted problem, V}",
      journal = {Astron. Astrophys},
         year = 1969,
        month = feb,
       volume = {1},
        pages = {223-238},
       adsurl = {https://ui.adsabs.harvard.edu/abs/1969A&A.....1..223H}
}

@ARTICLE{Henon70,
       author = {{Henon}, M.},
        title = "{Numerical exploration of the restricted problem. VI. Hill's case: Non-periodic orbits.}",
      journal = {Astron. Astrophys},
         year = 1970,
        month = nov,
       volume = {9},
        pages = {24-36},
       adsurl = {https://ui.adsabs.harvard.edu/abs/1970A&A.....9...24H}
}

@ARTICLE{Odstrcil03,
       author = {{Odstrcil}, D.},
        title = "{Modeling 3-D solar wind structure}",
      journal = {Advances in Space Research},
         year = 2003,
        month = aug,
       volume = {32},
       number = {4},
        pages = {497-506},
          doi = {10.1016/S0273-1177(03)00332-6},
       adsurl = {https://ui.adsabs.harvard.edu/abs/2003AdSpR..32..497O}
}

@ARTICLE{Perozzi17,
       author = {{Perozzi}, Ettore and {Ceccaroni}, Marta and {Valsecchi}, Giovanni B. and {Rossi}, Alessandro},
        title = "{Distant retrograde orbits and the asteroid hazard}",
      journal = {European Physical Journal Plus},
         year = 2017,
        month = aug,
       volume = {132},
       number = {8},
          eid = {367},
        pages = {367},
          doi = {10.1140/epjp/i2017-11644-0},
       adsurl = {https://ui.adsabs.harvard.edu/abs/2017EPJP..132..367P}
}

@ARTICLE{Pomoell18,
       author = {{Pomoell}, Jens and {Poedts}, S.},
        title = "{EUHFORIA: European heliospheric forecasting information asset}",
      journal = {Journal of Space Weather and Space Climate},
         year = 2018,
        month = jun,
       volume = {8},
          eid = {A35},
        pages = {A35},
          doi = {10.1051/swsc/2018020},
       adsurl = {https://ui.adsabs.harvard.edu/abs/2018JSWSC...8A..35P}
}

@INPROCEEDINGS{Provinciali24,
  author={Provinciali, Lorenzo and Calcagno, Davide and Amabili, Paride and Saita, Giorgio and Riccobono, Dario and Cicalò, Stefano and Marcucci, Maria Federica and Laurenza, Monica and Zimbardo, Gaetano and Landi, Simone and Walker, Roger},
  booktitle={2024 IEEE Aerospace Conference}, 
  title={HENON – Main Challenges of a Space Weather Alerts CubeSat Mission}, 
  year={2024},
  volume={},
  number={},
  pages={1-12},
  doi={10.1109/AERO58975.2024.10521299}}

@ARTICLE{Spencer11,
       author = {{Spencer}, E. and {Kasturi}, P. and {Patra}, S. and {Horton}, W. and {Mays}, M.~L.},
        title = "{Influence of solar wind-magnetosphere coupling functions on the Dst index}",
      journal = {Journal of Geophysical Research (Space Physics)},
         year = 2011,
        month = dec,
       volume = {116},
       number = {A12},
          eid = {A12235},
        pages = {A12235},
          doi = {10.1029/2011JA016780},
       adsurl = {https://ui.adsabs.harvard.edu/abs/2011JGRA..11612235S}
}

@ARTICLE{Thernisien06,
       author = {{Thernisien}, A.~F.~R. and {Howard}, R.~A. and {Vourlidas}, A.},
        title = "{Modeling of Flux Rope Coronal Mass Ejections}",
      journal = {Astrophys. J.},
         year = 2006,
        month = nov,
       volume = {652},
       number = {1},
        pages = {763-773},
          doi = {10.1086/508254},
       adsurl = {https://ui.adsabs.harvard.edu/abs/2006ApJ...652..763T}
}

@ARTICLE{Thernisien11,
       author = {{Thernisien}, A.},
        title = "{Implementation of the Graduated Cylindrical Shell Model for the Three-dimensional Reconstruction of Coronal Mass Ejections}",
      journal = {Astrophys. J. Suppl.},
         year = 2011,
        month = jun,
       volume = {194},
       number = {2},
          eid = {33},
        pages = {33},
          doi = {10.1088/0067-0049/194/2/33},
       adsurl = {https://ui.adsabs.harvard.edu/abs/2011ApJS..194...33T}
}

@ARTICLE{Verbanac13,
       author = {{Verbanac}, G. and {{\v{Z}}ivkovi{\'c}}, S. and {Vr{\v{s}}nak}, B. and {Bandi{\'c}}, M. and {Hojsak}, T.},
        title = "{Comparison of geoeffectiveness of coronal mass ejections and corotating interaction regions}",
      journal = {Astron. Astrophys.},
         year = 2013,
        month = oct,
       volume = {558},
          eid = {A85},
        pages = {A85},
          doi = {10.1051/0004-6361/201220417},
       adsurl = {https://ui.adsabs.harvard.edu/abs/2013A&A...558A..85V}
}

@ARTICLE{Vourlidas19,
       author = {{Vourlidas}, A. and {Patsourakos}, S. and {Savani}, N.~P.},
        title = "{Predicting the geoeffective properties of coronal mass ejections: current status, open issues and path forward}",
      journal = {Philosophical Transactions of the Royal Society of London Series A},
         year = 2019,
        month = jul,
       volume = {377},
       number = {2148},
        pages = {20180096},
          doi = {10.1098/rsta.2018.0096},
       adsurl = {https://ui.adsabs.harvard.edu/abs/2019RSPTA.37780096V}
}

@ARTICLE{Wang03,
       author = {{Wang}, Yuming and {Shen}, C.~L. and {Wang}, S. and {Ye}, P.~Z.},
        title = "{An empirical formula relating the geomagnetic storm's intensity to the interplanetary parameters: {$-VB_{z}$} and {$\Delta t$}}",
      journal = {Geophys. Res. Lett.},
         year = 2003,
        month = oct,
       volume = {30},
       number = {20},
          eid = {2039},
        pages = {2039},
          doi = {10.1029/2003GL017901},
       adsurl = {https://ui.adsabs.harvard.edu/abs/2003GeoRL..30.2039W}
}

@article{macalester2014extreme,
  title={Extreme space weather impact: An emergency management perspective},
  author={MacAlester, Mark H and Murtagh, William},
  journal={Space Weather},
  volume={12},
  number={8},
  pages={530--537},
  year={2014},
  doi = { 10.1002/2014SW001095 },
  publisher={Wiley Online Library}
}

@ARTICLE{Cicalo25,
       author = {{Cical{\`o}}, Stefano and {Alessi}, Elisa Maria and {Provinciali}, Lorenzo and {Amabili}, Paride and {Saita}, Giorgio and {Calcagno}, Davide and {Marcucci}, Maria Federica and {Laurenza}, Monica and {Zimbardo}, Gaetano and {Landi}, Simone and {Walker}, Roger and {Khan}, Michael},
        title = "{Mission analysis for the HENON CubeSat mission to a large Sun-Earth distant retrograde orbit}",
      journal = {Astrophys. Space Sci.},
         year = 2025,
        month = aug,
       volume = {370},
       number = {8},
          eid = {83},
        pages = {83},
          doi = {10.1007/s10509-025-04473-0},
archivePrefix = {arXiv},
       eprint = {2508.02138},
 primaryClass = {astro-ph.EP},
       adsurl = {https://ui.adsabs.harvard.edu/abs/2025Ap&SS.370...83C}
}

@article{gopalswamy2022SW,
  title={The Sun and space weather},
  author={Gopalswamy, Nat},
  journal={Atmosphere},
  volume={13},
  number={11},
  pages={1781},
  year={2022},
  doi = {10.3390/atmos13111781},
  publisher={MDPI}
}

@article{echer2005introduction,
  title={Introduction to space weather},
  author={Echer, E and Gonzalez, WD and Guarnieri, FL and Dal Lago, A and Vieira, LEA},
  journal={Advan. Space Res.},
  volume={35},
  number={5},
  pages={855--865},
  year={2005},
  doi = {10.1016/j.asr.2005.02.098},
  publisher={Elsevier}
}

@article{FRY2012,
title = {The risks and impacts of space weather: Policy recommendations and initiatives},
journal = {Space Policy},
volume = {28},
number = {3},
pages = {180-184},
year = {2012},
note = {Highlight: Assuring the sustainability of space activities},
issn = {0265-9646},
doi = {https://doi.org/10.1016/j.spacepol.2012.06.005},
url = {https://www.sciencedirect.com/science/article/pii/S0265964612000616},
author = {Emma Kiele Fry}
}

@article{kaiser2008stereo,
  title={The STEREO mission: An introduction},
  author={Kaiser, Michael L and Kucera, TA and Davila, JM and St. Cyr, OC and Guhathakurta, Madhulika and Christian, Eric},
  journal={Space Science Reviews},
  volume={136},
  number={1},
  pages={5--16},
  year={2008},
  doi = {10.1007/s11214-007-9277-0},
  publisher={Springer}
}

@article{brueckner1995large,
  title={The large angle spectroscopic coronagraph (LASCO) visible light coronal imaging and spectroscopy},
  author={Brueckner, GE and Howard, RA and Koomen, MJ and Korendyke, CM and Michels, DJ and Moses, JD and Socker, DG and Dere, KP and Lamy, PL and Llebaria, A and others},
  journal={Solar Physics},
  volume={162},
  number={1},
  pages={357--402},
  year={1995},
  doi = {10.1007/BF00733434},
  publisher={Springer}
}

@article{lemen2012atmospheric,
  title={The atmospheric imaging assembly (AIA) on the solar dynamics observatory (SDO)},
  author={Lemen, James R and Title, Alan M and Akin, David J and Boerner, Paul F and Chou, Catherine and Drake, Jerry F and Duncan, Dexter W and Edwards, Christopher G and Friedlaender, Frank M and Heyman, Gary F and others},
  journal={Solar Physics},
  volume={275},
  number={1},
  pages={17--40},
  year={2012},
  doi = {10.1007/s11207-011-9776-81},
  publisher={Springer}
}

@article{vourlidas2016wide,
  title={The wide-field imager for solar probe plus (WISPR)},
  author={Vourlidas, Angelos and Howard, Russell A and Plunkett, Simon P and Korendyke, Clarence M and Thernisien, Arnaud FR and Wang, Dennis and Rich, Nathan and Carter, Michael T and Chua, Damien H and Socker, Dennis G and others},
  journal={Space Science Reviews},
  volume={204},
  number={1},
  pages={83--130},
  year={2016},
  doi = { 
10.1007/s11214-014-0114-y },
  publisher={Springer}
}

@article{antonucci2020metis,
  title={Metis: the Solar Orbiter visible light and ultraviolet coronal imager},
  author={Antonucci, Ester and Romoli, Marco and Andretta, Vincenzo and Fineschi, Silvano and Heinzel, Petr and Moses, J Daniel and Naletto, Giampiero and Nicolini, Gianalfredo and Spadaro, Daniele and Teriaca, Luca and others},
  journal={Astronomy \& Astrophysics},
  volume={642},
  pages={A10},
  year={2020},
  doi = {10.1051/0004-6361/201935338},
  publisher={EDP Sciences}
}

@article{rodriguez2020space,
  title={Space weather monitor at the L5 point: A case study of a CME observed with STEREO B},
  author={Rodriguez, Luciano and Scolini, Camila and Mierla, Marilena and Zhukov, AN and West, MJ},
  journal={Space Weather},
  volume={18},
  number={10},
  pages={e2020SW002533},
  year={2020},
  doi = {10.1029/2020SW002533},
  publisher={Wiley Online Library}
}

@article{posner2021multi,
  title={A multi-purpose heliophysics L4 mission},
  author={Posner, Arik and Arge, Charles Nickolos and Staub, Jan and StCyr, Orville C and Folta, David and Solanki, Sami K and Strauss, Roelf Du Toit and Effenberger, Frederic and Gandorfer, Achim and Heber, Bernd and others},
  journal={Space Weather},
  volume={19},
  number={9},
  pages={e2021SW002777},
  year={2021},
  doi = {10.1002/essoar.10506845.1},
  publisher={Wiley Online Library}
}

@article{ravishankar2019estimation,
  title={Estimation of arrival time of coronal mass ejections in the vicinity of the Earth using SOlar and Heliospheric Observatory and Solar TErrestrial RElations Observatory observations},
  author={Ravishankar, Anitha and Micha{\l}ek, Grzegorz},
  journal={Solar Physics},
  volume={294},
  number={9},
  pages={125},
  year={2019},
  doi = {10.1007/s11207-019-1470-2},
  publisher={Springer}
}

@article{gopalswamy2011earth,
  title={Earth-Affecting Solar Causes Observatory (EASCO): a potential international living with a star mission from Sun--Earth L5},
  author={Gopalswamy, Natchimuthuk and Davila, Joseph M and Cyr, OC St and Sittler, Edward C and Auch{\`e}re, Frederic and Duvall Jr, TL and Hoeksema, JT and Maksimovic, Milan and MacDowall, RJ and Szabo, Adam and others},
  journal={Journal of Atmospheric and Solar-Terrestrial Physics},
  volume={73},
  number={5-6},
  pages={658--663},
  year={2011},
  doi = {10.1016/j.jastp.2011.01.013},
  publisher={Elsevier}
}

@article{webb2010using,
  title={Using STEREO-B as an L5 space weather pathfinder mission},
  author={Webb, DF and Biesecker, DA and Gopalswamy, N and Cyr, OC St and Davila, JM and Eyles, CJ and Thompson, BJ and Simunac, KDC and Johnston, JC},
  journal={Space Res. Today},
  volume={178},
  number={10},
  year={2010}
}

@ARTICLE{SOHO1995,
       author = {{Domingo}, V. and {Fleck}, B. and {Poland}, A.~I.},
        title = "{The SOHO Mission: an Overview}",
      journal = {\solphys},
         year = 1995,
        month = dec,
       volume = {162},
       number = {1-2},
        pages = {1-37},
          doi = {10.1007/BF00733425},
       adsurl = {https://ui.adsabs.harvard.edu/abs/1995SoPh..162....1D}
}

@ARTICLE{ACE1998,
       author = {{Smith}, C.~W. and {L'Heureux}, J. and {Ness}, N.~F. and {Acu{\~n}a}, M.~H. and {Burlaga}, L.~F. and {Scheifele}, J.},
        title = "{The ACE Magnetic Fields Experiment}",
      journal = {\ssr},
         year = 1998,
        month = jul,
       volume = {86},
        pages = {613-632},
          doi = {10.1023/A:1005092216668},
       adsurl = {https://ui.adsabs.harvard.edu/abs/1998SSRv...86..613S}
}

@ARTICLE{Wind1995,
       author = {{Acu{\~n}a}, M.~H. and {Ogilvie}, K.~W. and {Baker}, D.~N. and {Curtis}, S.~A. and {Fairfield}, D.~H. and {Mish}, W.~H.},
        title = "{The Global Geospace Science Program and Its Investigations}",
      journal = {\ssr},
         year = 1995,
        month = feb,
       volume = {71},
       number = {1-4},
        pages = {5-21},
          doi = {10.1007/BF00751323},
       adsurl = {https://ui.adsabs.harvard.edu/abs/1995SSRv...71....5A}
}

@ARTICLE{Scolini2019,
       author = {{Scolini}, C. and {Rodriguez}, L. and {Mierla}, M. and {Pomoell}, J. and {Poedts}, S.},
        title = "{Observation-based modelling of magnetised coronal mass ejections with EUHFORIA}",
      journal = {\aap},
         year = 2019,
        month = jun,
       volume = {626},
          eid = {A122},
        pages = {A122},
          doi = {10.1051/0004-6361/201935053},
archivePrefix = {arXiv},
       eprint = {1904.07059},
 primaryClass = {astro-ph.SR},
       adsurl = {https://ui.adsabs.harvard.edu/abs/2019A&A...626A.122S}
}

@ARTICLE{Chandrasekhar1957ApJ,
       author = {{Chandrasekhar}, S. and {Kendall}, P.~C.},
        title = "{On Force-Free Magnetic Fields.}",
      journal = {\apj},
         year = 1957,
        month = sep,
       volume = {126},
        pages = {457},
          doi = {10.1086/146413},
       adsurl = {https://ui.adsabs.harvard.edu/abs/1957ApJ...126..457C}
}

@article{Shiota2016,
author = {Shiota, D. and Kataoka, R.},
title = {Magnetohydrodynamic simulation of interplanetary propagation of multiple coronal mass ejections with internal magnetic flux rope (SUSANOO-CME)},
journal = {Space Weather},
volume = {14},
number = {2},
pages = {56-75},
doi = {https://doi.org/10.1002/2015SW001308},
url = {https://agupubs.onlinelibrary.wiley.com/doi/abs/10.1002/2015SW001308},
eprint = {https://agupubs.onlinelibrary.wiley.com/doi/pdf/10.1002/2015SW001308},
year = {2016}
}

@ARTICLE{verbeke2019A&A,
       author = {{Verbeke}, C. and {Pomoell}, J. and {Poedts}, S.},
        title = "{The evolution of coronal mass ejections in the inner heliosphere: Implementing the spheromak model with EUHFORIA}",
      journal = {\aap},
         year = 2019,
        month = jul,
       volume = {627},
          eid = {A111},
        pages = {A111},
          doi = {10.1051/0004-6361/201834702},
       adsurl = {https://ui.adsabs.harvard.edu/abs/2019A&A...627A.111V}
}

@ARTICLE{Maharana2022,
       author = {{Maharana}, Anwesha and {Isavnin}, Alexey and {Scolini}, Camilla and {Wijsen}, Nicolas and {Rodriguez}, Luciano and {Mierla}, Marilena and {Magdaleni{\'c}}, Jasmina and {Poedts}, Stefaan},
        title = "{Implementation and validation of the FRi3D flux rope model in EUHFORIA}",
      journal = {Advances in Space Research},
         year = 2022,
        month = sep,
       volume = {70},
       number = {6},
        pages = {1641-1662},
          doi = {10.1016/j.asr.2022.05.056},
archivePrefix = {arXiv},
       eprint = {2207.06707},
 primaryClass = {astro-ph.SR},
       adsurl = {https://ui.adsabs.harvard.edu/abs/2022AdSpR..70.1641M}
}

@article{Lundiquist1950,
author="Lundquist, S.",
title="Magnetohydrostatic fields",
journal="Ark. Fys.",
year="1950",
volume="2",
pages="361-365",
URL="https://cir.nii.ac.jp/crid/1570009749193268480"
}

@ARTICLE{Zimbardo23,
       author = {{Zimbardo}, G. and {Ying}, B. and {Nistic{\`o}}, G. and {Feng}, L. and {Rodr{\'\i}guez-Garc{\'\i}a}, L. and {Panasenco}, O. and {Andretta}, V. and {Banerjee}, D. and {Bemporad}, A. and {De Leo}, Y. and {Franci}, L. and {Frassati}, F. and {Habbal}, S. and {Long}, D. and {Magdalenic}, J. and {Mancuso}, S. and {Naletto}, G. and {Perri}, S. and {Romoli}, M. and {Spadaro}, D. and {Stangalini}, M. and {Strachan}, L. and {Susino}, R. and {Vainio}, R. and {Velli}, M. and {Cohen}, C.~M.~S. and {Giacalone}, J. and {Shen}, M. and {Telloni}, D. and {Abbo}, L. and {Burtovoi}, A. and {Jerse}, G. and {Landini}, F. and {Nicolini}, G. and {Pancrazzi}, M. and {Russano}, G. and {Sasso}, C. and {Uslenghi}, M.},
        title = "{A high-latitude coronal mass ejection observed by a constellation of coronagraphs: Solar Orbiter/Metis, STEREO-A/COR2, and SOHO/LASCO}",
      journal = {\aap},
         year = 2023,
        month = aug,
       volume = {676},
          eid = {A48},
        pages = {A48},
          doi = {10.1051/0004-6361/202346011},
       adsurl = {https://ui.adsabs.harvard.edu/abs/2023A&A...676A..48Z}
}

@ARTICLE{Iyer2006JApA,
       author = {{Iyer}, K.~N. and {Jadav}, R.~M. and {Jadeja}, A.~K. and {Manoharan}, P.~K. and {Sharma}, Som and {Vats}, Hari Om},
        title = "{Space Weather Effects of Coronal Mass Ejection}",
      journal = {Journal of Astrophysics and Astronomy},
         year = 2006,
        month = jun,
       volume = {27},
       number = {2-3},
        pages = {219-226},
          doi = {10.1007/BF02702524},
       adsurl = {https://ui.adsabs.harvard.edu/abs/2006JApA...27..219I}
}

@inproceedings{Hill2003,
author = {Steve Hill and Victor J. Pizzo},
title = {{Advanced solar imaging from the GOES R spacecraft}},
volume = {4853},
booktitle = {Innovative Telescopes and Instrumentation for Solar Astrophysics},
editor = {Stephen L. Keil and Sergey V. Avakyan},
organization = {International Society for Optics and Photonics},
publisher = {SPIE},
pages = {465 -- 478},
year = {2003},
doi = {10.1117/12.460385},
URL = {https://doi.org/10.1117/12.460385}
}

@ARTICLE{Lotaniu2023,
       author = {{Loto'aniu}, Paul T.~M. and {Davis}, A. and {Jarvis}, A. and {Grotenhuis}, M. and {Rich}, F.~J. and {Califf}, S. and {Inceoglu}, F. and {Pacini}, A. and {Singer}, H.~J.},
        title = "{Initial on-Orbit Results from the GOES-18 Spacecraft Science Magnetometer}",
      journal = {\ssr},
         year = 2023,
        month = dec,
       volume = {219},
       number = {8},
          eid = {84},
        pages = {84},
          doi = {10.1007/s11214-023-01032-3},
       adsurl = {https://ui.adsabs.harvard.edu/abs/2023SSRv..219...84L}
}

@ARTICLE{Stiefel2025A&A,
       author = {{Stiefel}, Muriel Zo{\"e} and {Kuhar}, Matej and {Limousin}, Olivier and {Dickson}, Ewan C.~M. and {Volpara}, Anna and {Hurford}, Gordon J. and {Krucker}, S{\"a}m},
        title = "{Using the STIX background detector as a proxy for GOES}",
      journal = {\aap},
         year = 2025,
        month = feb,
       volume = {694},
          eid = {A138},
        pages = {A138},
          doi = {10.1051/0004-6361/202452574},
archivePrefix = {arXiv},
       eprint = {2501.03667},
 primaryClass = {astro-ph.SR},
       adsurl = {https://ui.adsabs.harvard.edu/abs/2025A&A...694A.138S}
}

@ARTICLE{Fox2016,
       author = {{Fox}, N.~J. and {Velli}, M.~C. and {Bale}, S.~D. and {Decker}, R. and {Driesman}, A. and {Howard}, R.~A. and {Kasper}, J.~C. and {Kinnison}, J. and {Kusterer}, M. and {Lario}, D. and {Lockwood}, M.~K. and {McComas}, D.~J. and {Raouafi}, N.~E. and {Szabo}, A.},
        title = "{The Solar Probe Plus Mission: Humanity's First Visit to Our Star}",
      journal = {\ssr},
         year = 2016,
        month = dec,
       volume = {204},
       number = {1-4},
        pages = {7-48},
          doi = {10.1007/s11214-015-0211-6},
       adsurl = {https://ui.adsabs.harvard.edu/abs/2016SSRv..204....7F}
}

@ARTICLE{Kaiser2005AdSpR,
       author = {{Kaiser}, M.~L.},
        title = "{The STEREO mission: an overview}",
      journal = {Advances in Space Research},
         year = 2005,
        month = jan,
       volume = {36},
       number = {8},
        pages = {1483-1488},
          doi = {10.1016/j.asr.2004.12.066},
       adsurl = {https://ui.adsabs.harvard.edu/abs/2005AdSpR..36.1483K}
}

@ARTICLE{muller2020A&A,
       author = {{M{\"u}ller}, D. and {St. Cyr}, O.~C. and {Zouganelis}, I. and {Gilbert}, H.~R. and {Marsden}, R. and {Nieves-Chinchilla}, T. and {Antonucci}, E. and {Auch{\`e}re}, F. and {Berghmans}, D. and {Horbury}, T.~S. and {Howard}, R.~A. and {Krucker}, S. and {Maksimovic}, M. and {Owen}, C.~J. and {Rochus}, P. and {Rodriguez-Pacheco}, J. and {Romoli}, M. and {Solanki}, S.~K. and {Bruno}, R. and {Carlsson}, M. and {Fludra}, A. and {Harra}, L. and {Hassler}, D.~M. and {Livi}, S. and {Louarn}, P. and {Peter}, H. and {Sch{\"u}hle}, U. and {Teriaca}, L. and {del Toro Iniesta}, J.~C. and {Wimmer-Schweingruber}, R.~F. and {Marsch}, E. and {Velli}, M. and {De Groof}, A. and {Walsh}, A. and {Williams}, D.},
        title = "{The Solar Orbiter mission. Science overview}",
      journal = {\aap},
         year = 2020,
        month = oct,
       volume = {642},
          eid = {A1},
        pages = {A1},
          doi = {10.1051/0004-6361/202038467},
archivePrefix = {arXiv},
       eprint = {2009.00861},
 primaryClass = {astro-ph.SR},
       adsurl = {https://ui.adsabs.harvard.edu/abs/2020A&A...642A...1M}
}

@ARTICLE{Sangalli25,
       author = {{Sangalli}, M. and {Verdini}, A. and {Landi}, S. and {Papini}, E.},
        title = "{The effects of expansion and turbulence on the interplanetary evolution of a magnetic cloud}",
      journal = {\aap},
         year = 2025,
        month = jul,
       volume = {699},
          eid = {A258},
        pages = {A258},
          doi = {10.1051/0004-6361/202554559},
archivePrefix = {arXiv},
       eprint = {2505.15527},
 primaryClass = {astro-ph.SR},
       adsurl = {https://ui.adsabs.harvard.edu/abs/2025A&A...699A.258S}
}

@INPROCEEDINGS{Arge2003AIPC,
       author = {{Arge}, Charles N. and {Odstrcil}, Dusan and {Pizzo}, Victor J. and {Mayer}, Leslie R.},
        title = "{Improved Method for Specifying Solar Wind Speed Near the Sun}",
    booktitle = {Solar Wind Ten},
         year = 2003,
       editor = {{Velli}, Marco and {Bruno}, Roberto and {Malara}, Francesco and {Bucci}, B.},
       series = {American Institute of Physics Conference Series},
       volume = {679},
        month = sep,
    publisher = {AIP},
        pages = {190-193},
          doi = {10.1063/1.1618574},
       adsurl = {https://ui.adsabs.harvard.edu/abs/2003AIPC..679..190A}
}

@ARTICLE{Xie2004JGRA,
       author = {{Xie}, Hong and {Ofman}, Leon and {Lawrence}, Gareth},
        title = "{Cone model for halo CMEs: Application to space weather forecasting}",
      journal = {Journal of Geophysical Research (Space Physics)},
         year = 2004,
        month = mar,
       volume = {109},
       number = {A3},
          eid = {A03109},
        pages = {A03109},
          doi = {10.1029/2003JA010226},
       adsurl = {https://ui.adsabs.harvard.edu/abs/2004JGRA..109.3109X}
}

@ARTICLE{Kataoka2009JGRA,
       author = {{Kataoka}, R. and {Ebisuzaki}, T. and {Kusano}, K. and {Shiota}, D. and {Inoue}, S. and {Yamamoto}, T.~T. and {Tokumaru}, M.},
        title = "{Three-dimensional MHD modeling of the solar wind structures associated with 13 December 2006 coronal mass ejection}",
      journal = {Journal of Geophysical Research (Space Physics)},
         year = 2009,
        month = oct,
       volume = {114},
       number = {A10},
          eid = {A10102},
        pages = {A10102},
          doi = {10.1029/2009JA014167},
       adsurl = {https://ui.adsabs.harvard.edu/abs/2009JGRA..11410102K}
}

@article{Isavnin_2016,
doi = {10.3847/1538-4357/833/2/267},
url = {https://dx.doi.org/10.3847/1538-4357/833/2/267},
year = {2016},
month = {dec},
publisher = {The American Astronomical Society},
volume = {833},
number = {2},
pages = {267},
author = {Isavnin, A.},
title = {FRiED: A NOVEL THREE-DIMENSIONAL MODEL OF CORONAL MASS EJECTIONS},
journal = {The Astrophysical Journal}
}

@ARTICLE{Mayaud1980GMS,
       author = {{Mayaud}, P.~N.},
        title = "{Derivation, Meaning, and Use of Geomagnetic Indices}",
      journal = {Geophysical Monograph Series},
         year = 1980,
        month = jan,
       volume = {22},
        pages = {607},
          doi = {10.1029/GM022},
       adsurl = {https://ui.adsabs.harvard.edu/abs/1980GMS....22..607M}
}

@INPROCEEDINGS{Latiff2024,
       author = {{Latiff}, Zatul Iffah Abd and {Hairuddin}, Muhammad Asraf and {Zainuddin}, Aznilinda and {Ashar}, Nur Dalila Khirul and {Jusoh}, Mohamad Huzaimy},
        title = "{Analytical Approach to SYM-H based Geomagnetic Storm Classifications using Statistical Features Extraction}",
    booktitle = {Journal of Physics Conference Series},
         year = 2024,
       series = {Journal of Physics Conference Series},
       volume = {2915},
        month = dec,
    publisher = {IOP},
          eid = {012010},
        pages = {012010},
          doi = {10.1088/1742-6596/2915/1/012010},
       adsurl = {https://ui.adsabs.harvard.edu/abs/2024JPhCS2915a2010L}
}

@article{Gonzalez1987,
title = {Criteria of interplanetary parameters causing intense magnetic storms (Dst < −100 nT)},
journal = {Planetary and Space Science},
volume = {35},
number = {9},
pages = {1101-1109},
year = {1987},
issn = {0032-0633},
doi = {https://doi.org/10.1016/0032-0633(87)90015-8},
url = {https://www.sciencedirect.com/science/article/pii/0032063387900158},
author = {Walter D. Gonzalez and Bruce T. Tsurutani},
}

@article{wang2004deflection,
  title={Deflection of coronal mass ejection in the interplanetary medium},
  author={Wang, Yuming and Shen, Chenglong and Wang, S and Ye, Pinzhong},
  journal={Solar Physics},
  volume={222},
  number={2},
  pages={329--343},
  year={2004},
  doi = {10.1023/B:SOLA.0000043576.21942.aa },
  publisher={Springer}
}

@ARTICLE{2024Mayank,
       author = {{Mayank}, Prateek and {Lotz}, Stefan and {Vaidya}, Bhargav and {Mishra}, Wageesh and {Chakrabarty}, D.},
        title = "{Study of Evolution and Geo-effectiveness of Coronal Mass Ejection{\textendash}Coronal Mass Ejection Interactions Using Magnetohydrodynamic Simulations with SWASTi Framework}",
      journal = {\apj},
         year = 2024,
        month = nov,
       volume = {976},
       number = {1},
          eid = {126},
        pages = {126},
          doi = {10.3847/1538-4357/ad8084},
archivePrefix = {arXiv},
       eprint = {2409.19943},
 primaryClass = {astro-ph.SR},
       adsurl = {https://ui.adsabs.harvard.edu/abs/2024ApJ...976..126M}
}

@ARTICLE{altschuler1969magnetic,
       author = {{Altschuler}, Martin D. and {Newkirk}, Jr., Gordon},
        title = "{Magnetic Fields and the Structure of the Solar Corona. I: Methods of Calculating Coronal Fields}",
      journal = {\solphys},
         year = 1969,
        month = sep,
       volume = {9},
       number = {1},
        pages = {131-149},
          doi = {10.1007/BF00145734},
       adsurl = {https://ui.adsabs.harvard.edu/abs/1969SoPh....9..131A}
}

@article{schatten1969model,
  title={A model of interplanetary and coronal magnetic fields},
  author={Schatten, Kenneth H and Wilcox, John M and Ness, Norman F},
  journal={Solar Physics},
  volume={6},
  number={3},
  pages={442--455},
  year={1969},
  doi = {10.1007/BF00146478},
  publisher={Springer}
}

@article{deforest2026polarimeter,
  title={Polarimeter to Unify the Corona and Heliosphere (PUNCH)},
  author={DeForest, Craig E and Gibson, Sarah E and Killough, Ronnie and Waltham, Nick R and Beasley, Matt N and Colaninno, Robin C and Laurent, Glenn T and Seaton, Daniel B and Hughes, J Marcus and Guhathakurta, Madhulika and others},
  journal={Solar Physics},
  volume={301},
  number={1},
  pages={16},
  year={2026},
  doi={https://doi.org/10.1007/s11207-026-02608-2},
  publisher={Springer}
}

@inproceedings{maruca2022radial,
  title={Radial Trends in Plasma Parameters Across the Heliosphere},
  author={Maruca, Bennett and Qudsi, Ramiz A and Alterman, BLL and Walsh, Brian and Korreck, Kelly E and Verscharen, Daniel and Bandyopadhyay, Riddhi and Chhiber, Rohit and Chasapis, Alexandros and Parashar, Tulasi and others},
  booktitle={AGU Fall Meeting Abstracts},
  volume={2022},
  pages={SH35D--1837},
  year={2022}
}

@article{ruiz2014characterization,
  title={Characterization of the turbulent magnetic integral length in the solar wind: from 0.3 to 5 astronomical units},
  author={Ruiz, Maria Emilia and Dasso, S and Matthaeus, WH and Weygand, JM},
  journal={Solar Physics},
  volume={289},
  number={10},
  pages={3917--3933},
  year={2014},
  publisher={Springer}
}

@article{cuesta2022isotropization,
  title={Isotropization and evolution of energy-containing eddies in solar wind turbulence: Parker solar probe, helios 1, ace, wind, and voyager 1},
  author={Cuesta, Manuel Enrique and Chhiber, Rohit and Roy, Sohom and Goodwill, Joshua and Pecora, Francesco and Jarosik, Jake and Matthaeus, William H and Parashar, Tulasi N and Bandyopadhyay, Riddhi},
  journal={The Astrophysical Journal Letters},
  volume={932},
  number={1},
  pages={L11},
  year={2022},
  doi={10.3847/2041-8213/ac73fd},
  publisher={IOP Publishing}
}
   
%\end{linenumbers}

\end{document}